\documentclass[journal]{IEEEtran}
\usepackage{cite}
\usepackage{amsmath,amssymb,amsfonts}
\usepackage{algorithmic}
\usepackage{graphicx}
\usepackage{textcomp}
\usepackage{authblk}
\usepackage{xcolor}
\usepackage{makecell}
\usepackage{float}
\usepackage[hyphens]{url}
\usepackage{subcaption}
\usepackage{tablefootnote}
\usepackage{pgfplots}
\usepackage{multirow}
\usepackage{listings}
\usepackage{courier}
\usepackage[bookmarks=true,breaklinks=true,colorlinks,citecolor=blue,linkcolor=blue,urlcolor=blue]{hyperref}
\usepackage{cleveref}

\pgfplotsset{width=7cm,compat=1.3}
\usetikzlibrary{pgfplots.groupplots, positioning,decorations.pathreplacing}
\definecolor{orange}{cmyk}{0,0.5,1,0}

\definecolor{codegreen}{rgb}{0,0.6,0}
\definecolor{codegray}{rgb}{0.5,0.5,0.5}
\definecolor{codepurple}{rgb}{0.58,0,0.82}
\definecolor{backcolour}{rgb}{0.95,0.95,0.92}
\lstdefinestyle{mystyle}{
    backgroundcolor=\color{backcolour},   
    commentstyle=\color{codegreen},
    keywordstyle=\color{magenta},
    numberstyle=\tiny\color{codegray},
    stringstyle=\color{codepurple},
    basicstyle=\footnotesize\ttfamily,
    breakatwhitespace=false,         
    breaklines=true,                 
    captionpos=b,                    
    keepspaces=true,                 
    numbers=left,                    
    numbersep=5pt,                  
    showspaces=false,                
    showstringspaces=false,
    showtabs=false,                  
    tabsize=2
}

\definecolor{bblue}{HTML}{4F81BD}
\definecolor{rred}{HTML}{C0504D}
\definecolor{ggreen}{HTML}{9BBB59}
\definecolor{ppurple}{HTML}{9F4C7C}

\usepackage{tikz}
\usetikzlibrary{shapes.geometric, arrows, decorations.markings, arrows.meta, backgrounds, shadows, fit}
\tikzset{%
  >={Latex[width=2mm,length=2mm]},
            base/.style = {rectangle, rounded corners, draw=black,
                           minimum width=4cm, minimum height=1cm,
                           text centered, font=\sffamily},
  phase1/.style = {base, fill=blue!30},
  phase2/.style = {base, fill=red!30},
  phase3/.style = {base, fill=green!30},
  evaluate/.style = {base, minimum width=2.5cm, fill=orange!15,
                           font=\ttfamily},
  client/.style = {rectangle, rounded corners, draw=black,
                           minimum width=4cm, minimum height=1cm,
                           text centered, fill=yellow!30},
  fpga/.style = {rectangle, rounded corners, draw=black,
                           minimum width=2cm, minimum height=1cm,
                           text centered, fill=red!30},
  electronics/.style = {rectangle, rounded corners, draw=black,
                           minimum width=2cm, minimum height=1cm,
                           text centered, fill=orange!30,    general shadow = {%
      shadow scale = 1,
      shadow xshift = -1ex,
      shadow yshift = 1ex,
      draw,
      thick,
      fill = orange!30},
    general shadow = {%
      shadow scale = 1,
      shadow xshift = -0.5ex,
      shadow yshift = 0.5ex,
      draw,
      thick,
      fill = orange!30},}, 
 frontend/.style = {rectangle, rounded corners, draw=black,
                           minimum width=4cm, minimum height=1cm,
                           text centered, fill=blue!30},
  ir/.style = {rectangle, rounded corners, draw=black,
                           minimum width=2cm, minimum height=1cm,
                           text centered, fill=red!30},
  backend/.style = {rectangle, rounded corners, draw=black,
                           minimum width=4cm, minimum height=1cm,
                           text centered, fill=green!30}
}
\tikzstyle{vecArrow} = [thick, decoration={markings,mark=at position
   1 with {\arrow[semithick]{open triangle 60}}},
   double distance=1.4pt, shorten >= 5.5pt,
   preaction = {decorate},
   postaction = {draw,line width=1.4pt, white,shorten >= 4.5pt}]

\def\BibTeX{{\rm B\kern-.05em{\sc i\kern-.025em b}\kern-.08em
    T\kern-.1667em\lower.7ex\hbox{E}\kern-.125emX}}

\newcommand{\TP}[1]{}
\newcommand{\term}[1]{``{#1}''}
\title{A Classical Architecture For Digital Quantum Computers}
\author[1]{Fang Zhang$^*$}
\author[2]{Xing Zhu$^*$}
\author[1]{Rui Chao}
\author[1]{Cupjin Huang}
\author[2]{Linghang Kong}
\author[4]{Guoyang Chen}
\author[3]{Dawei Ding}
\author[5]{Haishan Feng}
\author[2]{Yihuai Gao}
\author[2]{Xiaotong Ni}
\author[2]{Liwei Qiu}
\author[5]{Zhe Wei}
\author[5]{Yueming Yang}
\author[5]{Yang Zhao}
\author[1]{Yaoyun Shi}
\author[4]{Weifeng Zhang}
\author[4]{Peng Zhou}
\author[1]{Jianxin Chen}
\affil[1]{Quantum Laboratory, DAMO Academy, Bellevue, Washington 98004, USA}
\affil[2]{Quantum Laboratory, DAMO Academy, Hangzhou, Zhejiang 311121, P.R.China}
\affil[3]{Quantum Laboratory, DAMO Academy, Sunnyvale, California 94085, USA}
\affil[4]{Alibaba Cloud Intelligence, Alibaba Group USA, Sunnyvale, California 94085, USA}
\affil[5]{Alibaba Cloud Intelligence, Alibaba Group, Hangzhou, Zhejiang 311121, P.R.China}

\date{}                     
\begin{document}
\maketitle
\def\thefootnote{*}\footnotetext{These authors contributed equally to this work.}\def\thefootnote{\arabic{footnote}}
\thispagestyle{plain}
\pagestyle{plain}


\begin{abstract}

Scaling bottlenecks the making of digital quantum computers, posing challenges from both the quantum and the classical components. 
We present a classical architecture to cope with a comprehensive list of the latter challenges {\em all at once}, and implement it fully in an end-to-end system by integrating a multi-core \mbox{RISC-V} CPU with our in-house control electronics. 

Our architecture enables scalable, high-precision control of large quantum processors and accommodates evolving requirements of quantum hardware. A central feature is a microarchitecture executing quantum operations in parallel on arbitrary predefined qubit groups. Another key feature is a reconfigurable quantum instruction set that supports easy qubit re-grouping and instructions extensions.

As a demonstration, we implement the widely-studied surface code quantum computing workflow, which is instructive for being demanding on both the controllers and the integrated classical computation. Our design, for the first time, reduces instruction issuing and transmission costs to constants, which do not scale with the number of qubits, without adding any overheads in decoding or dispatching.

Rather than relying on specialized hardware for syndrome decoding, our system uses a dedicated general-purpose multi-core CPU for both qubit control and classical computation, including syndrome decoding. This simplifies the system design and facilitates load-balancing between the quantum and classical components. We implement recent theoretical proposals as decoding firmware on a \mbox{RISC-V} system-on-chip that parallelizes general inner decoders. By using various inner decoders, including our in-house Union-Find and PyMatching 2 implementations, we can achieve unprecedented decoding capabilities of up to distances 47 and 67 with the currently available systems-on-chips (SoCs), under realistic and optimistic assumptions of physical error rate $p=0.001$ and $p=0.0001$, respectively, all in just 1 \textmu s.
\end{abstract}

\section{Motivations and Summary of Results}

As quantum computers become more sophisticated~\cite{AAB+19, egan2021fault, WBC+21, google2021exponential}, their demands on the classical control multiply accordingly. In this section, we analyze those challenges, then summarize our solutions. We confine this work to the superconducting-circuit platform, the focus of our team. We first review the setup as the starting point for our discussion.

\begin{figure}[ht]
\centering
\includegraphics[trim=1mm 1mm 1mm 1mm,clip,scale=0.5]{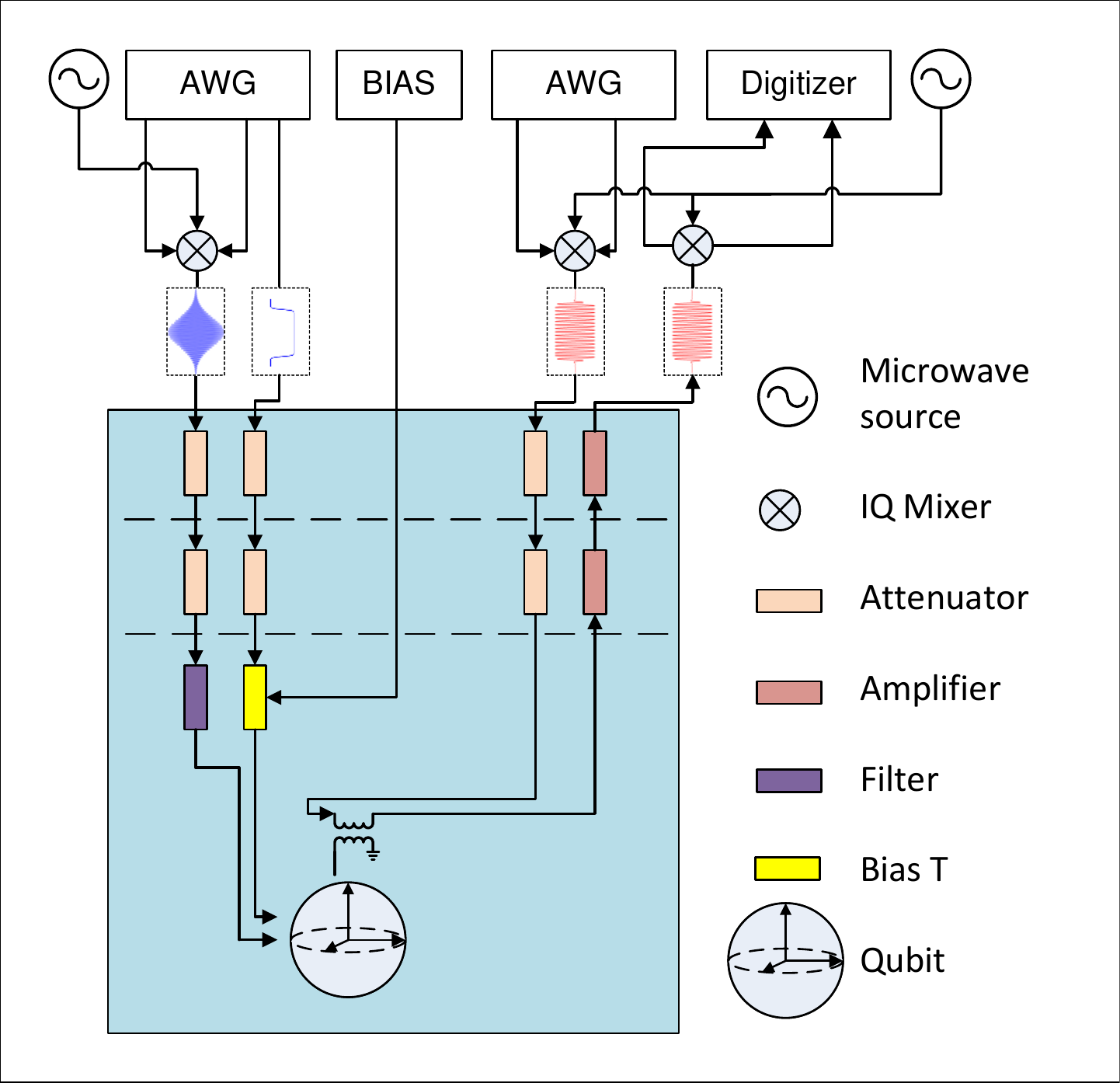}
\caption{An experimental setup for qubit driving and  measurement. The dilution refrigerator is depicted as the cyan box, with different temperature zones separated by dashed lines. the PC driving the control electronics is omitted.}
\label{fig:setup}
\end{figure}

\textbf{Superconducting system setup.} \Cref{fig:setup} illustrates a standard setup for a superconducting quantum computing system. Quantum information is stored physically on superconducting qubits on a quantum chip. To enable superconductivity and to suppress thermal noise, the quantum chip is cooled cryogenically inside a dilution refrigerator. To enable state evolution and measurement, the superconducting circuits are coupled to drive lines connecting to room-temperature control electronics, which in turn comprise of arbitrary waveform generators (AWGs), digitizers, IQ mixers, etc. The control electronics are further driven by a general-purpose processing unit such as a PC.

\textbf{Quantum computing workflows.}
Applications are the end goals of quantum computers, thus the origins of their design requirements. Most applications belong to one of the two main paradigms: noisy intermediate-scale quantum (NISQ) applications and fault-tolerant quantum computations (FTQC). NISQ applications operate on noisy, unprotected physical qubits, limited in scale and in precision. FTQCs operate on encoded logical qubits, each consisting of (likely) thousands of physical qubits. The logical qubits have drastically reduced sensitivity to physical-level noises, allowing computations of an arbitrary length and scale, thus consequently the ultimate quantum advantages.

In NISQ, the PC sends the quantum circuit to the control electronics. The latter parse the circuit into microwave waveforms, play them synchronously on the drive lines to the qubits, process the measurement responses from the quantum chip, and finally, return the measurement results to the PC. The PC can then perform a classical post-processing, before possibly starting the next round of quantum circuit execution. 

FTQC differs from NISQ in several key aspects. First, it requires constant extraction and decoding of the classical error syndromes, which are constantly churned out by the faulty quantum circuits. The decoding in turn requires real-time and intense classical computation. Second, while NISQ executes a static circuit, FTQC requires dynamic quantum circuit generation according to the decoding results. 

Both NISQ and FTQC demand seamless coordination and collaboration between classical and quantum computational resources, which in turn require a co-design of classical and quantum architecture. We focus on the design and implementation of classical architectures. We analyze the challenges from two perspectives: scaling-up and actual implementation of a complete system.

\textbf{Challenges in scaling up the classical architecture.}
Maintaining a high precision in the control of quantum hardware is the primary requirement here as it would directly affect fidelities of the quantum operations involved. Failing it would result in performance loss that eventually needs to be compensated by the quantum hardware, compounding the difficulty for the latter.
Specifically, for superconducting qubits, microwave pulses played on different AWG channels and the sampling window of digitizer channels need to be synchronized at the picosecond level to ensure high-fidelity physical operations\cite{yang2022fpga}. 

A second set of challenges are caused by the large number of instructions --- the efficiency of their issuance, transmission, and execution as the number of qubits grows. These problems have been recognized by several authors~\cite{FRR+19,CGR20,BMW+20}, and we refer to them together as \term{instruction stresses}.
In FTQC, dynamic quantum instructions need to be issued and transmitted fast enough to keep in pace with the rapid quantum execution, posing a hard constraint on the classical architecture. This may not be required for NISQ, but is still desirable as it would decrease the total running time. 

Syndrome decoding is yet another major bottleneck to FTQC classical architecture~\cite{terhal2015quantum}. For surface code schemes on present-day superconducting qubits, one round of syndrome extraction takes roughly 1\textmu s~\cite{googleqec}, and generates $O(d^2)$ bits of syndrome information in parallel, for $d$ being the code distance. Against this increasing syndrome size, the 
decoding algorithm needs to keep up with the constant syndrome extraction time and in order to avoid exponential syndrome backlog.

Multiple decoding schemes were proposed to tackle this problem~\cite{Fow15,DPM+20,HJP+20,UKT+21,DLJ22}, but can only handle code distances no more than $11$, even with specialized hardware. Recently, a new parallel decoding scheme was proposed independently in~\cite{TZC+22,SBB+22}. An implementation of the scheme achieved a code distance of $11$ for physical error rate $p=0.4\%$~\cite{Riv22}. 

A fourth set of challenges originate from a desirable feature that we call \term{permissiveness}, which means the ability to accommodate evolving requirements by other components of a quantum computer. Our field experiences indicate that implementing a complete classical architecture is time-consuming and labor-intensive. On the other hand, in this early stage of quantum computing, changes are rapid in applications, hardware characteristics, and error-correction schemes. Thus a stable yet permissive classical architecture would be cost-effective in the classical-quantum co-design process. 

\textbf{Challenges for implementing a complete system.}
Many researchers have proposed innovative solutions addressing one or a few of the above problems. Ultimately, a single system needs to be built for a real quantum computer. Building such a complete system has the additional challenge of balancing competing objectives with currently available and compatible technologies. To our knowledge, there has not been a system implementation addressing all the aforementioned challenges in scalability.

\textbf{Our contributions.} 
We present and implement a classical architecture to address all the scalability challenges mentioned above in one single system. 

\begin{enumerate}
\item Our system provides \textit{high-fidelity qubit control} by interconnecting one-chassis PXIe systems through a star-like hierarchy with high-density connectors. This design synchronizes, with high accuracy, pulses from different control electronics, enabling precise qubit control even as the system size increases, thereby maintaining high-fidelity. This conclusion is supported by the extensive testing of the channel-to-channel and phase jitter of the AWG outputs.
\item To address \textit{instruction stresses}, we develop an efficient \term{quantum instruction pipeline} that combines Single-Instruction-Multiple-Data (SIMD) with a broadcasting mechanism. This pipeline enables parallel application of the same type of gate on arbitrarily-sized qubit groups. The operation types and the qubit groups of application-specific instructions can be easily configured either prior to the execution of the quantum program or during runtime. Additionally, the costs across instruction issuing, transmission,
dispatching and execution does not scale with the size of each qubit group.
\item Our system's \textit{permissiveness} is achieved through a combination of features, including a reconfigurable quantum instruction set, Memory-Mapped IO (MMIO) in the microarchitecture, and a portable general-purpose CPU. The instruction set and the underlying MMIO-based microarchitecture facilitate the incorporation of new quantum instructions. 
\item We achieve unprecedented performances on \textit{decoding throughput} for surface codes, a mainstream approach that our architecture and the implemented system are nevertheless not restricted to. More specifically, we implement the surface code and a parallel decoding firmware based on the recent theoretical proposals~\cite{TZC+22, Riv22} in a dedicated CPU and benchmark its performance on a development board. By leveraging our in-house
Union-Find and PyMatching 2\cite{pymatchingv2} as inner decoders, we can decode up to distances $13$ and $31$ on SiFive P650\cite{SiFive22} or T-head C910\cite{CXL+20}, or $41$ and $67$
with ET-SoC-1\cite{DEA+21}, all in just 1 microsecond for physical error rate $p=0.0001$.
\end{enumerate}

The proposed classical architecture is implemented fully in an end-to-end quantum computer system by integrating a multi-core, vectorized \mbox{RISC-V} CPU with our in-house control electronics. Our system also features an MLIR-based compiler to support the proposed reconfigurable quantum instruction set and enables optimization possibilities on various abstraction layers.
We highlight the following features among the many of our implemented system.

\begin{enumerate}
    \item[5)] \textit{Low communication latency} A key metric for FTQC is the \term{decoding latency}, i.e. the time between the completions of syndrome generation and decoding. Such latency consists of the decoding algorithm latency and the communication latency between the control system and the quantum device. In our design, we use on-board communication to reduce the latter. This design also enables other capabilities where latency plays a critical role, such as on-the-fly calibration\cite{navarathna2021neural,machado2019recent,granade2015accelerated}, and just-in-time compilation\cite{wallman2016noise,wilson2020just}.
    \item[6)] \textit{Load balancing through multi-core CPU} The bottleneck in classical computation is not always syndrome decoding, and can vary during the computational process. To accommodate different scenarios, we use a dedicated multi-core CPU in our system, allowing dynamic allocation of cores to syndrome decoding, qubit control or other computation-heavy tasks. This design allows us to achieve optimal performance while avoiding the unnecessary complexities and cost of using specialized hardware for syndrome decoding.
\end{enumerate}

\textbf{Comparison with previous work.}\label{sec:comparison}  

To the best of our knowledge, our architecture proposal and the resulting actual implementation represent the first attempt to address, in a single system, the above comprehensive list of scaling challenges for the classical architecture.

Instruction stresses have been known for long, with various mitigating approaches proposed~\cite{FRR+19, CGR20, BMW+20}. Those include using Single-Instruction-Multiple-Data (SIMD) and Very-Long-Instruction-Word (VLIW) to reduce the instruction issuance rate~\cite{FRR+19}, and multiprocessors to increase quantum operation and circuit-level parallelism~\cite{ZXZ+21}. However, those methods provide only constant-factor improvements and are insufficient to cope with the increasing overhead that scales with the code distance in surface code quantum computing.

The QuEST proposal~\cite{TMN+17} addresses the instruction bandwidth problem by employing dedicated programmable micro-code engines. While it shows promises for enabling real-time instruction issuing, it crucially relies on an assumption from the underlying primeline microarchitecture~\cite{hornibrook2015cryogenic}: that all qubits driven at a given time must share the same frequency.

This may hold for some quantum computing platforms, such as cold atoms or trapped ions, but not for superconducting qubits, where frequency differences are likely inevitable and sometimes a design preference. 
Furthermore, the absence of scaling analysis makes it unclear how the frequency requirement would affect the performance in an actual implementation.

Syndrome decoding has been another well-known concern in the quantum computing community for over a decade\cite{terhal2015quantum}, with proposals ranging from efficient algorithms to specific microarchitectures~\cite{delfosse2021almost, DLJ22, pymatchingv2, DPM+22}. Before our work, it remained an open problem if a general-purpose CPU with on-board communication to the control electronics would be  sufficient to provide the required decoding throughput. We answer this question affirmatively for the first time by combining the recent parallel decoding schemes~\cite{TZC+22, Riv22} with an efficient in-house implementation for the Union-Find decoder and a recent implementation of the Minimum Weight Perfect Matching (MWPM) algorithm~\cite{pymatchingv2}.

\begin{figure*}[ht]
\centering
\includegraphics[trim=1mm 1mm 1mm 1mm,clip,scale=0.28]{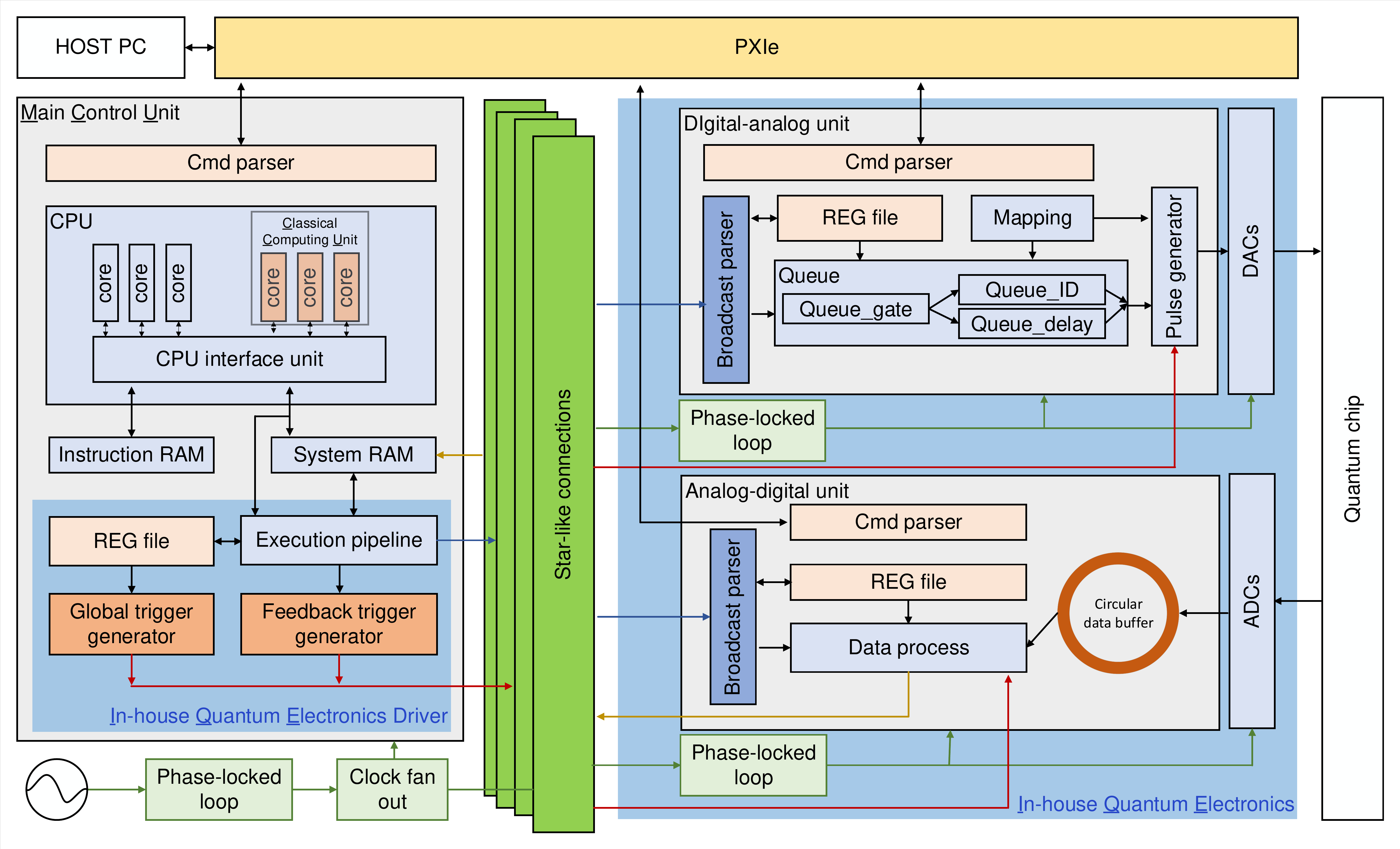}
\caption{Block diagram of the proposed classical architecture. The architecture consists of a host PC, a main control unit (MCU), control electronics (In-house Quantum Electronics). The quantum chip is connected with the control electronics via drive lines, while the host PC, the MCU and the electronics are jointly connected via PXIe. Additionally, the MCU connects with all electronics via a star-like connection. A dedicated unit in the MCU is responsible for driving the control electronics. The MCU is equipped with a portable CPU, and a portion of it, called the classical computing unit, is allocated for run-time computation-heavy tasks such as syndrome decoding. In our implementation, the digital-analog and analog-digital units are made in-house and are called in-house quantum electronics (IQE), and their corresponding driver unit in the MCU is called the IQE driver. Please note that all other modules can be configured via the command parser; however, we have omitted the corresponding arrows in the diagram for the sake of simplicity.
}
\label{fig:micro}
\end{figure*}

\section{Architecture Design and System Implementation}\label{sec:architecture}

\subsection{Architecture Design}
\label{subsec:arc}
\TP{Summary description of the Master FPGA}
See~\Cref{fig:micro} for a block diagram of our system design. The MCU contains a dedicated CPU. Besides controlling the qubits via the electronics driver, the CPU can also execute classical tasks offloaded from the host PC, using dedicated cores labeled the classical computing unit (CCU). Such tasks naturally arise from logical quantum program execution, dynamic calibration, and just-in-time compilation, etc. The offloading greatly shortens the communication latency with the QPU.

A quantum program generally comprises both quantum and classical components that collaborate to solve a problem. In \Cref{sec:evaluation}, we will introduce our front-end language and the corresponding compilation support. However, the design and workflow of our system are not restricted to specific quantum programming languages. When a quantum program is executed on a host PC, the quantum subroutines and, depending on the implementation, potentially some classical subroutines will be sent to the MCU. The MCU then processes these quantum or quantum-classical hybrid tasks by issuing both classical and quantum instructions. Classical instructions are carried out on the dedicated CPU for classical control and computations, while quantum instructions are executed through requests to our in-house quantum electronics (IQE) driver.
The IQE driver dispatches corresponding \term{IQE instructions}, or \term{commands}, to IQE, which in turn drives the quantum processor. At the CPU level, the \term{quantum instructions} are implemented as pseudo-instructions that expand to MMIO load/store instructions. These MMIO instructions interact with a special memory region, and the electronics driver decodes them and dispatches \term{electronics-level instructions}, which will be explained shortly, through broadcasting for communication with the control electronics.

The electronics-level instructions specify the pulse sequences and their corresponding timing information to the control electronics. The latter parse the instructions and feed the pulse sequence information into a local queue. The pulse sequence is not played until a special \term{trigger} signal arrives at the control electronics, which then plays the pulse sequence through its ports and empties the queue, waiting for the next round of pulse instructions.

\begin{figure*}
    \centering
    \includegraphics[width=0.85\textwidth, trim=1cm 0 1cm 0, clip]{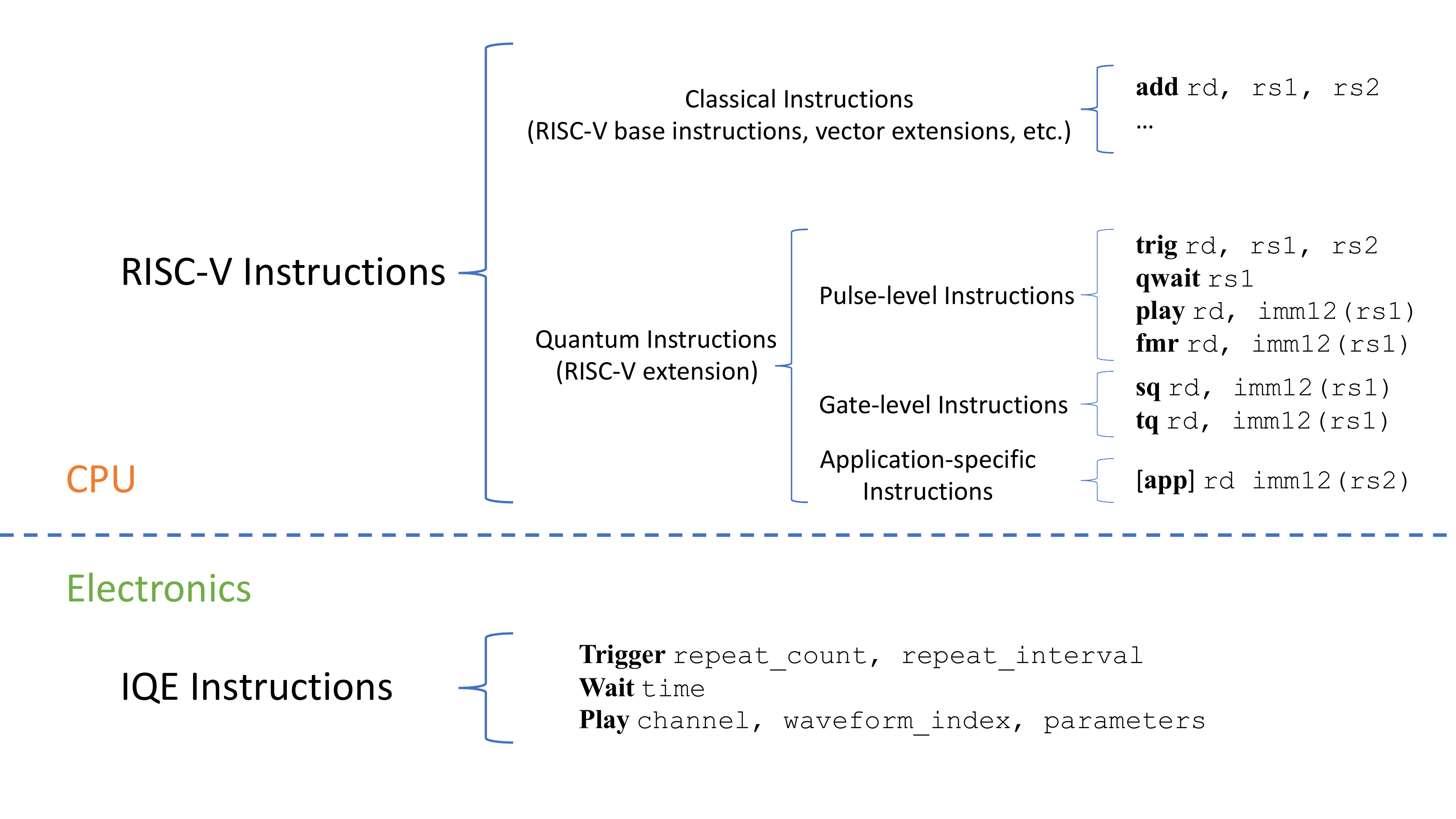}
\caption{Taxonomy of instructions.}
    \label{fig:instructions}
\end{figure*}

We use various \term{instruction} terminologies. For clarity, \Cref{fig:instructions} exhibits a taxonomy, with more details in the main text. 

In addition to the aforementioned general setup, a key feature of our architecture is a \textit{quantum instruction pipeline} that naturally supports a large number of parallel repetition of a same gate, and allows for easy reconfiguration. This is enabled jointly by several components, which we elaborate below.

\textbf{Reconfigurable quantum instruction set.}
Exploiting MMIO's flexibility, our modular quantum instruction set comprises of \term{pulse-level instructions} for qubit control and calibration, and \term{gate-level instructions} for quantum circuit execution. By having both levels available, it allows for flexibility in implementing quantum algorithms and calibrating quantum devices, similar to other systems~\cite{FRR+19,ZXZ+21,BMW+20}. We distinctly exploit what we call the {\em brickwork structure} found in typical quantum circuits: there is a small number of single-layer sub-circuit of the form $\bigotimes_i G^{S_i}$, for some partition $\{S_i\}_i$ of either the whole set or a large subset of the qubits into equal-sized subsets, and an identical gate $G$ acting on each subset. We allocate different MMIO addresses for the partition identifiers, and specify the gate type via the message written to the address.  Decoding and dispatching of the instruction are left to the underlying microarchitecture. This allows a lightweight specification of application-specific instructions on user-defined qubit partitions, which in turn significantly alleviates the cost of instruction issuing and data transmission.

\textbf{Instruction dispatching via broadcasting.} Some designs may prioritize certain aspects of instruction processing at the expense of others. For instance, adding complex instructions to reduce the instruction issuing rate can lead to more complex decoding and dispatching. However, our microarchitecture support does not come with any hidden costs. This means that we have successfully reduced costs at every stage of the instruction processing pipeline. When dispatching a single gate instruction to multiple control electronics, one-to-one communication would scale the cost linearly with the number of control electronics, impeding scalability. We avoid this problem by exploiting the few-distinct-partition property of the brickwork structure through the built-in broadcasting mechanism of the star-like connection. 

Each signal transmitted from the electronics driver broadcasts automatically through the star-like connection, thus each IQE instruction is sent to a collection of control electronics simultaneously, regardless if a control electronic is meant to be involved in the instruction. To utilize this, each electronic device holds a \term{partition mask} specifying which partitions it is in. Each IQE instruction broadcast from the electronics driver comes with a partition identifier. Upon receiving an instruction, each electronic device checks whether the partition identifier of the broadcast instruction matches one of the partition identifiers in its partition mask. If so, it proceeds to process the instruction, and ignores it otherwise.

The MCU thus can issue at once a same instruction to all devices with a common partition identifier, realizing microarchitecture-level single-instruction-multiple-destination (SIMD). The partition masks are stored in local registration entry (REG) files on each electronic device. They are easily reconfigurable, either using PXIe between runs or in real-time via the same star-like connection.

\textbf{Instruction decoding.}
To decode a quantum instruction received from the CPU, the electronics driver extracts the electronics-level instruction type determined by the value written through MMIO, and appends it with the partition identifier determined by the MMIO address. Both mappings are stored in a local REG file that can be reconfigured if necessary. The assembled electronics-level instruction is then dispatched through the broadcasting system mentioned before. 

This microarchitecture does not incur extra overhead on either decoding or dispatching when the partition size increases (as in the case of more qubits). 

Apart from the above quantum instruction pipeline problems, we highlight some design choices that address scaling.

\textbf{Pulse synchronization via triggers.} 
All of the ADCs and the DACs are driven in the same clock domain through a phase-locked loop and a star-like connection, with one rubidium oscillator used as the system root clock.
Our design further synchronizes pulses on different control electronics via a dedicated trigger mechanism. The pulses are not played through DACs immediately upon processing of the electronics-level instructions, but rather are stored in a local queue. When a trigger instruction is issued from the MCU, the trigger signal arrives at each control electronic device at the same time, guaranteeing pulse-level synchronization. For further information on IQE, please refer to \cite{yang2022fpga}.

\textbf{Portable, tightly integrated but loosely coupled dedicated CPU.} Unlike previous schemes~\cite{FRR+19,CGR20,BMW+20,ZXZ+21} that handle the communication of the control electronics and the MCU by new and dedicated CPU instructions, ours aims to avoid substantial CPU modifications thus works solely with the unmodified classical instruction set instead. The MCU and the electronics driver are coupled only through MMIO instructions. This loose coupling provides portability and extensibility, as the same communication scheme can in principle be used with all CPUs supporting the same underlying ISA, or even other classical ISAs, with little to no modification. On the other hand, the dedicated CPU is tightly integrated to the control electronics through onboard communication which significantly reduce the communication cost.

\subsection{System implementation}
Our design can in principle be implemented over various classical and quantum hardware. For our particular implementation, we assume room-temperature, as opposed to cryogenic, electronics as they are more widely deployed today. While there is no fundamental reason to prefer \mbox{RISC-V}, ARM, or other instruction set architectures, we choose \mbox{RISC-V} for its potential in future system evolution. For instance, we anticipate that integrating the required quantum instruction pipelines into the \mbox{RISC-V} IP core would be less limited due to its open license business model.

\textbf{Hardware setup.} 
We implement a prototype by integrating a \mbox{RISC-V} IP core with our room-temperature electronics, which include a timing control module (TCM), four-channel AWGs, four-channel data acquisition modules, a local oscillator, amplifiers, mixers, and a high-precision voltage source. As mentioned above, the in-house AWGs and the digitaizers are collectively referred to as IQE. 

We implement a real-time digital signal processing system on built-in FPGAs of the IQE, featuring precise timing control, arbitrary waveform generation, and parallel IQ demodulation for qubit state discrimination. The FPGA in TCM serves as the master FPGA running the MCU and the IQE driver. The master FPGA communicates with the AWGs and the digitizers through high-speed digital backplane transmissions.

\begin{figure}[ht]
\centering
\includegraphics[trim=1mm 1mm 1mm 1mm,clip,scale=0.5]{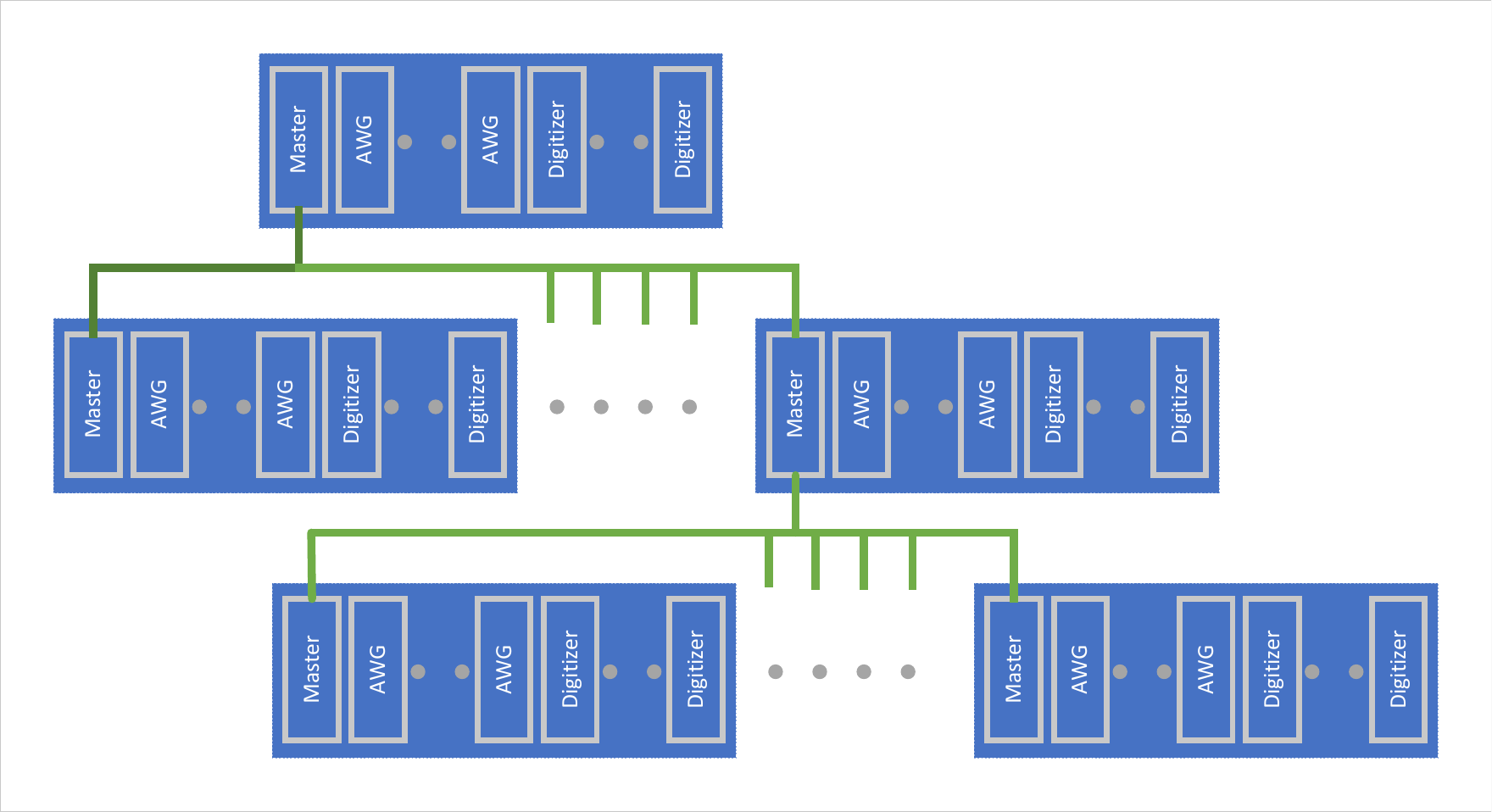}
\caption{Expansion scheme via star-like connection.}
\label{fig:expand_hierarchy}
\end{figure}

In the aforementioned configuration depicted in \Cref{fig:micro}, we have assumed an unrestricted number of connections in the star-like topology. Now, we will explain how we can scale up from chassis-based systems that have a limited number of connections. A standard chassis with $18$ slots meet the requirements of $10$ qubits' control and readout. In such a one-chassis PXIe system, the master FPGA with a soft \mbox{RISC-V} IP core is used to provide triggers and instructions to other AWGs and digitizers. In order to control more qubits, the master FPGA in each one-chassis PXIe system can be interconnected through high-density connectors via a star-like expansion, as illustrated in~\Cref{fig:expand_hierarchy}. Only one master FPGA needs to implement the soft \mbox{RISC-V} IP core as the MCU of the whole system. The MCU broadcasts the instructions to the master FPGAs of all those one-chassis PXIe subsystems by a daisy chain interface or star-like interface, and then each master FPGA broadcasts to AWGs and digitizers in the same chassis.

We will now provide a comprehensive overview of various types of instructions present in our architecture, including the IQE instructions  as well as the RISC-V quantum instructions. Together they facilitate seamless control over the quantum processor. 

\TP{IQE instructions}
\textbf{IQE instructions.} 
We specify the electronic-level instructions, or commands, broadcast by the IQE driver via the star-like connection, hereafter referred to as the \term{IQE instructions} for convenience (note that those \term{instructions} are not directly related to any CPU-level instructions). Currently, there are three types of IQE instructions, as summarized in \Cref{tab:IQE_instructions}:
\begin{itemize}
    \item 
\textit{``Trigger''}: As mentioned above, the ``Trigger'' instruction tells the IQE driver to actually start executing all quantum operations in the queue. To facilitate repeated measurements frequently occurring in qubit calibration, we ensure that the IQE driver has  built-in functionality to \emph{repeat} all quantum operations in the queue, with a specified number of repetitions and time intervals.
    \item\textit{``Wait''}: The ``Wait'' instruction controls the relative timing between quantum operations in the same trigger, giving the user program full control on scheduling.
    \item\textit{``Play''}: The ``Play'' instruction is the most basic quantum instruction. It plays a predefined waveform or a predefined combination of waveforms on one or more channels, which corresponds to quantum operations like qubit reset, 1- or 2-qubit gates, or qubit measurement under the computational bases. Quantum measurement instructions differ from other operations in that they yield a result; these two cases are differentiated by the corresponding waveform indices, where indices $i\ge 128$ correspond to measurements, while indices $< 128$ are for no return values. The digitizer decodes the measurement instructions and sets the sampling window according to the parameters specified in the instruction. After IQ channel demodulation and data processing, the digitizer transmits the result to the CPU's system RAM.

\end{itemize}
\begin{table}[ht]
\centering
\begin{tabular}{l|l}
Type & Operands \\
\hline
Trigger & \textbf{Repeat count}, \textit{repeat interval} \\
Wait & \textbf{Time} \\
Play & \underline{Channel}, \textbf{waveform index}, \textit{parameters} 
\end{tabular}
\caption{Summary of instructions to the IQE driver. In the \term{Operands} column, underlined text indicates the \term{address operand} (an operand that is determined by the memory address rather than a value written); bold text indicates the \term{main operand} (i.e., writing this operand issues the instruction to the IQE driver); all other operands are in italic text, indicating that writing to them does not issue the instruction and that their values are preserved in the IQE driver memory.}
\label{tab:IQE_instructions}
\end{table}

\textbf{Quantum instruction set.}
We here present an instantiation of a modular set of pseudo-instructions at the CPU level, consisting of pulse-level instructions for device calibration and gate-level instructions for algorithm implementations. These pseudo-instructions are not implemented directly, but are subsequently expanded to \mbox{RISC-V} MMIO operations via built-in load/store instructions. We also provide an example MMIO layout compatible with existing \mbox{RISC-V} architectures.

\TP{Overview of the instruction set}
\Cref{tab:RISC-V_instructions} illustrates the current design of the quantum instruction set and its corresponding expansion into MMIO load/store instructions. The instruction set features a hierarchical design, consisting of pulse-level, gate-level, and application-specific instructions. Each higher-level instruction can be decomposed into lower-level instructions with the same functionality, but using higher-level instructions reduced the decoding and dispatching overhead.
\begin{itemize}
\item \term{Pulse-level Instructions} \texttt{play}, \texttt{qwait} and \texttt{trig}: specify pulses, their relative timing, and the issuance of the trigger signal, respectively. More precisely, \texttt{play} specifies the actual control pulse sequence, \texttt{qwait} specifies the scheduling of the corresponding pulses, and \texttt{trig} triggers the actual execution of previously issued instructions. Additionally, \texttt{fmr} loads the qubit measurement results from previous runs from the predetermined addresses. 
    \item \term{Gate-level Instructions} \texttt{sq} and \texttt{tq}: correspond to single-qubit and two-qubit operations, respectively. They share a similar expansion as the pulse-level `play' instructions, with different address offsets. As a single-qubit operation (e.g., a gate, a measurement, the qubit reset) usually consists of pulse plays on different physical channels with different timing constraints, a gate-level instruction is usually decoded into multiple IQE instructions by the IQE driver. This enables a relatively decoupled design of the MCU and the quantum architecture, as the exact interpretation of gate operations is only defined at the IQE driver level. 
    \item \term{Application-specific Instructions} \texttt{app}:  shares the same instruction format as the pulse-level \texttt{play}, but its decoding into IQE instructions is entirely left to the user. The user can design the decoding of \texttt{app} instructions as different combinations of IQE instructions, as long as it does not incur too much overhead on the IQE driver side. As the use cases of quantum processors in the near and far future remains largely uncertain, such customizable instruction design provides freedom of exploration with different potential use cases, including NISQ and fault-tolerant quantum computation. 

\end{itemize}
We show in \Cref{tab:mmio_specification} an example MMIO layout supporting 0x4000, or $16,384$, physical qubits, compatible with existing \mbox{RISC-V} architecture designs, including Si-Five, T-head, etc. 
The allocated address space supports individual qubit control over quantum memory experiments on a surface code patch with a code distance of $90$, or lattice surgery on two qubits with a code distance of up to $64$. Note that neither distances is a hard constraint as the MMIO address space is easily extendible to support a larger-scale computation.

\begin{table*}[ht]
\centering
\scalebox{0.9}{
\begin{tabular}{c|l|l|l}
Pseudo-instruction & Base instruction(s) & Meaning of parameters\tablefootnote{The load/store instructions \texttt{lw, sw, sb} in this column are used to transfer a `word' or a `byte' between memory and registers.}\\
\hline
\multirow{7}{*}{Pulse-level}&\texttt{trig rd, rs1, rs2} & \makecell[l]{\texttt{sw rd, ADDR\char`_TRIGGER+8}\\\texttt{sw rs1, ADDR\char`_TRIGGER+4}\\\texttt{sw rs2, ADDR\char`_TRIGGER}} & \makecell[l]{\texttt{rd} --- bit mask of channels\\\texttt{rs1} --- repeat count\\\texttt{rs2} --- repeat interval} \\
\cline{2-4}
&\texttt{qwait rs1} & \texttt{sw rs1, ADDR\char`_WAIT} & \texttt{rs} --- time \\
\cline{2-4}
&\texttt{play rd, imm12(rs1)} & \makecell[l]{\texttt{sb rd, ADDR\char`_PLAY + imm12(rs1)}} & \makecell[l]{\texttt{rd} --- waveform index\\\texttt{imm12(rs1)} --- memory offset for the channel} \\
\cline{2-4}
&\texttt{fmr rd, imm12(rs1)} & \texttt{lw rd, ADDR\char`_FMR + imm12(rs1)} & \makecell[l]{\texttt{rd} --- destination register\\\texttt{imm12(rs1)} --- \parbox[t]{0.22\textwidth}{\raggedright result storage address}}\\
\hline
\hline
\multirow{3}{*}{Gate-level}&\texttt{sq rd, imm12(rs1)} & \makecell[l]{\texttt{sb rd, ADDR\char`_GATE1Q + imm12(rs1)}} & \makecell[l]{\texttt{rd} --- gate index\\\texttt{imm12(rs1)} --- memory offset for the qubit} \\
\cline{2-4}
&\texttt{tq rd, imm12(rs1)} & \makecell[l]{\texttt{sb rd, ADDR\char`_GATE2Q + imm12(rs1)}} & \makecell[l]{\texttt{rd} --- gate index\\\texttt{imm12(rs1)} --- memory offset for the qubit pair} \\
\hline\hline
Application-specific&
\texttt{app rd, imm12(rs1)} & \makecell[l]{\texttt{sb rd, ADDR\char`_APP + imm12(rs1)}} & \makecell[l]{\texttt{rd} --- operation index\\\texttt{imm12(rs1)} --- \parbox[t]{0.22\textwidth}{\raggedright memory offset for\\ the operation grouping}}  \\
\end{tabular}}
\caption{Summary of \mbox{RISC-V} pseudo-instructions designed for communicating with the AQE driver. \texttt{ADDR\char`_*} are memory addresses that are determined at design time and thus are constant in the assembler.}
\label{tab:RISC-V_instructions}
\end{table*}

\begin{table}[ht]
\centering
\begin{tabular}{l|l|l}
Name & Address & Type \\
\hline
\texttt{ADDR\char`_TRIGGER} & 0x40001000 & int32 \\
\texttt{ADDR\char`_WAIT} & 0x40002000 & int32 \\
\texttt{ADDR\char`_FMR} & 0x40003000 & int32[0x1400] \\
\texttt{ADDR\char`_SQ} & 0x40010000 & uint8[0x4000] \\
\texttt{ADDR\char`_TQ} & 0x40014000 & uint8[0x8000] \\
\texttt{ADDR\char`_PLAY} & 0x4001c000 & uint8[0x8000] \\
\texttt{ADDR\char`_APP} & 0x40024000 & uint8[0x4000] \\
\end{tabular}
\caption{An example of MMIO address layout.}
\label{tab:mmio_specification}
\end{table}

\begin{figure*}[ht]
    \centering
    {\includegraphics[width=0.8\linewidth, trim={0 0.38cm 6cm 1cm},clip]{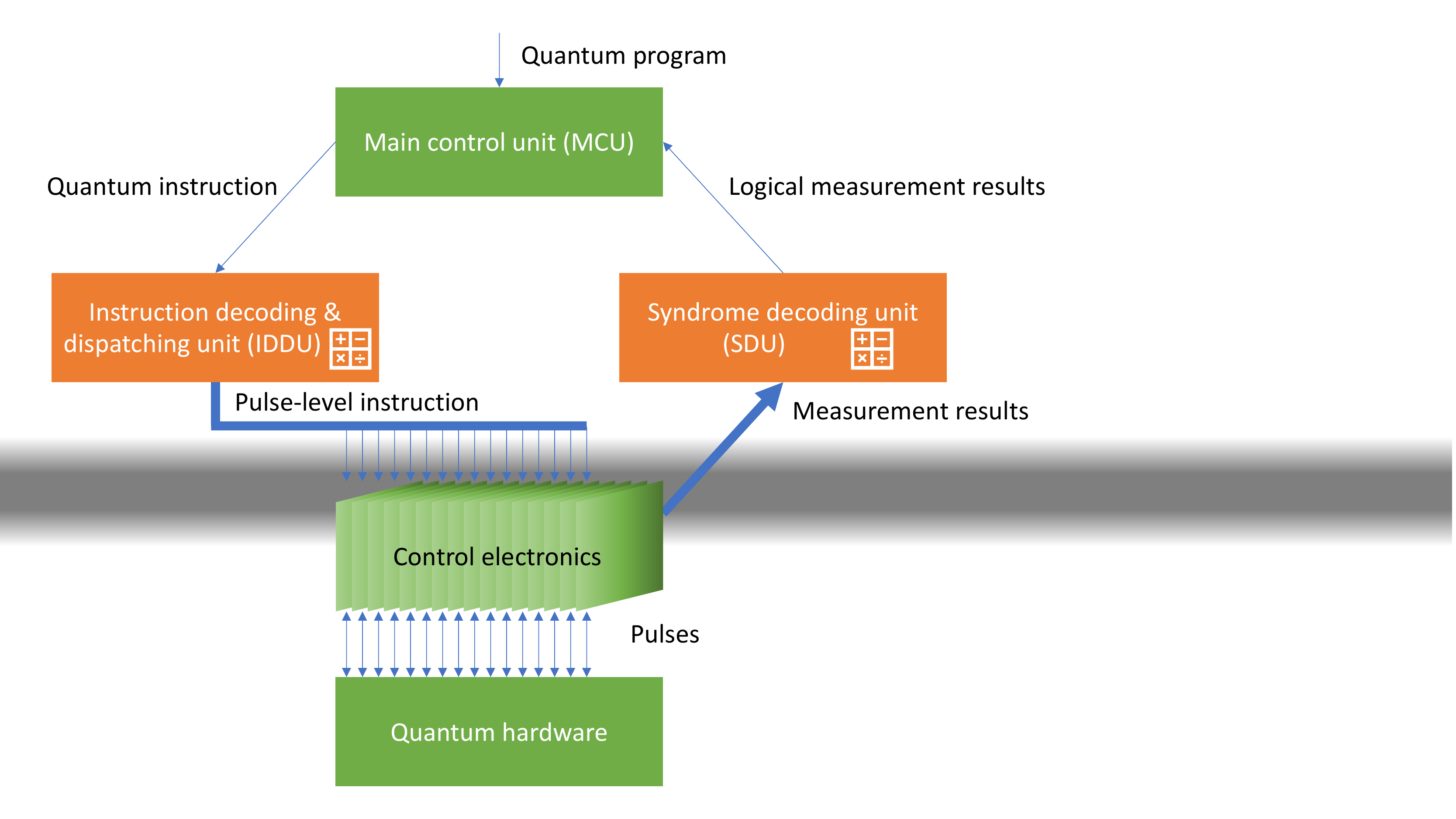}}
    \caption{Schematic workflow supporting surface code quantum computation (SCQC). The shaded area illustrates the blurred boundary of ``classical'' and ``quantum'' architecture, the former being our main focus.}
    \label{fig:workflow}
\end{figure*}

\section{Evaluation methodology}

A full demonstration of scalability requires a large-scale quantum processor yet to be built. We thus focus on implementing essential features over surface code quantum computing using our architecture, to argue that known scalability challenges can be resolved through our design. 

More specifically, we reach our conclusion by focusing on components involved in large-scale computation or communication, and analyzing how the incurred costs scale with the code distance and the quality of the quantum device.
Our architecture design is not specific to surface-code-based quantum computing, thus can be readily generalized to other quantum error correcting codes or fault-tolerant schemes.

\subsection{Surface code quantum computation}

\TP{Intro to surface code}

Surface code encodes the logical information of one qubit into a patch of $d\times d$ physical qubits, such that any error happening on at most $\lfloor (d-1)/2\rfloor$ physical qubits can be detected through intermediate measurements and be corrected accordingly. A popular approach to realize logical Clifford operations for the surface code is \term{lattice surgery}~\cite{HFD+12}. 
Specifically, patches of logical qubits are arranged on a large grid, with additional physical qubits positioned in the \term{routing space}~\cite{CC22} between them. 
Then, lattice surgery allows measuring logical Pauli jointly over multiple patches, using interactions only between pairs of nearest-neighbor physical qubits.

There are alternatives to lattice surgery for realizing logical operations on surface codes (see~\cite{bombin2023logical} and references therein). In this work, by \term{surface code quantum computation} (SCQC), we refer to the approach through lattice surgery.

\TP{FTQC workflow overview}
\Cref{fig:workflow} shows a schematic workflow of SCQC from the perspective of classical control. Upon receiving a pre-compiled quantum program, the MCU issues quantum instructions to an \term{instruction decoding and dispatching unit} (IDDU) through MMIO when needed. The IDDU then decodes the quantum instructions into pulse-level instructions readily executable on each of the electronics and dispatches them accordingly. The control electronics interact with the quantum hardware and return the raw measurement results to a dedicated memory region. 

The incoming syndrome information is fed to and decoded by a classical firmware, called the \term{syndrome decoding unit} (SDU), that runs on the dedicated CPU.
Once decoded, the logical measurement results are fed to the MCU for adaptive real-time generation of the future instructions required by fault-tolerant quantum computing. In our implementation,  the IDDU, the electronics and the SDU correspond respectively to the IQE driver, the IQE and part of the CCU.

\begin{figure}[ht]
\centering
    \begin{subfigure}{0.3\linewidth}
    \centering
    \includegraphics[height=3.5cm]{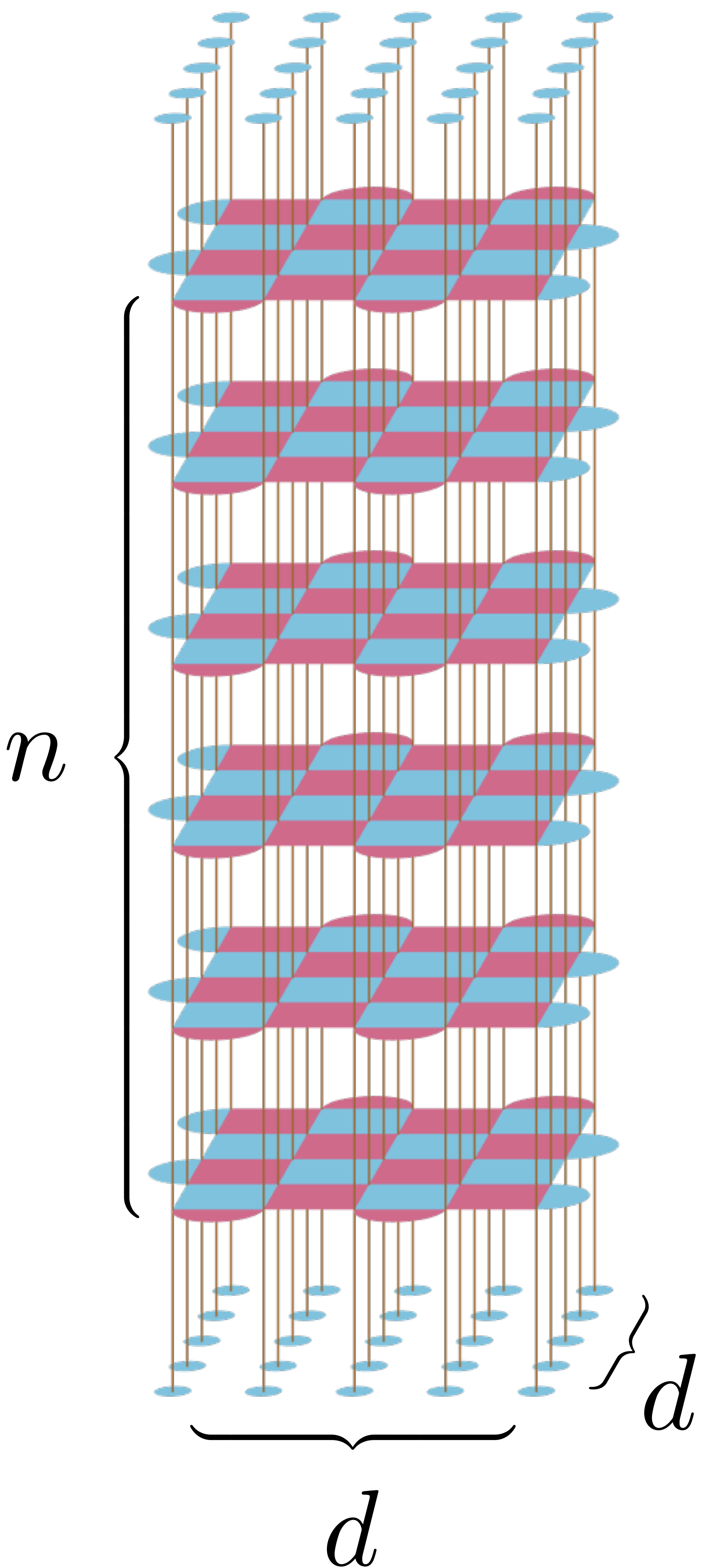}
        \caption{}
    \end{subfigure}
    \begin{subfigure}{0.6\linewidth}
    \centering
    \includegraphics[height=3.5cm]{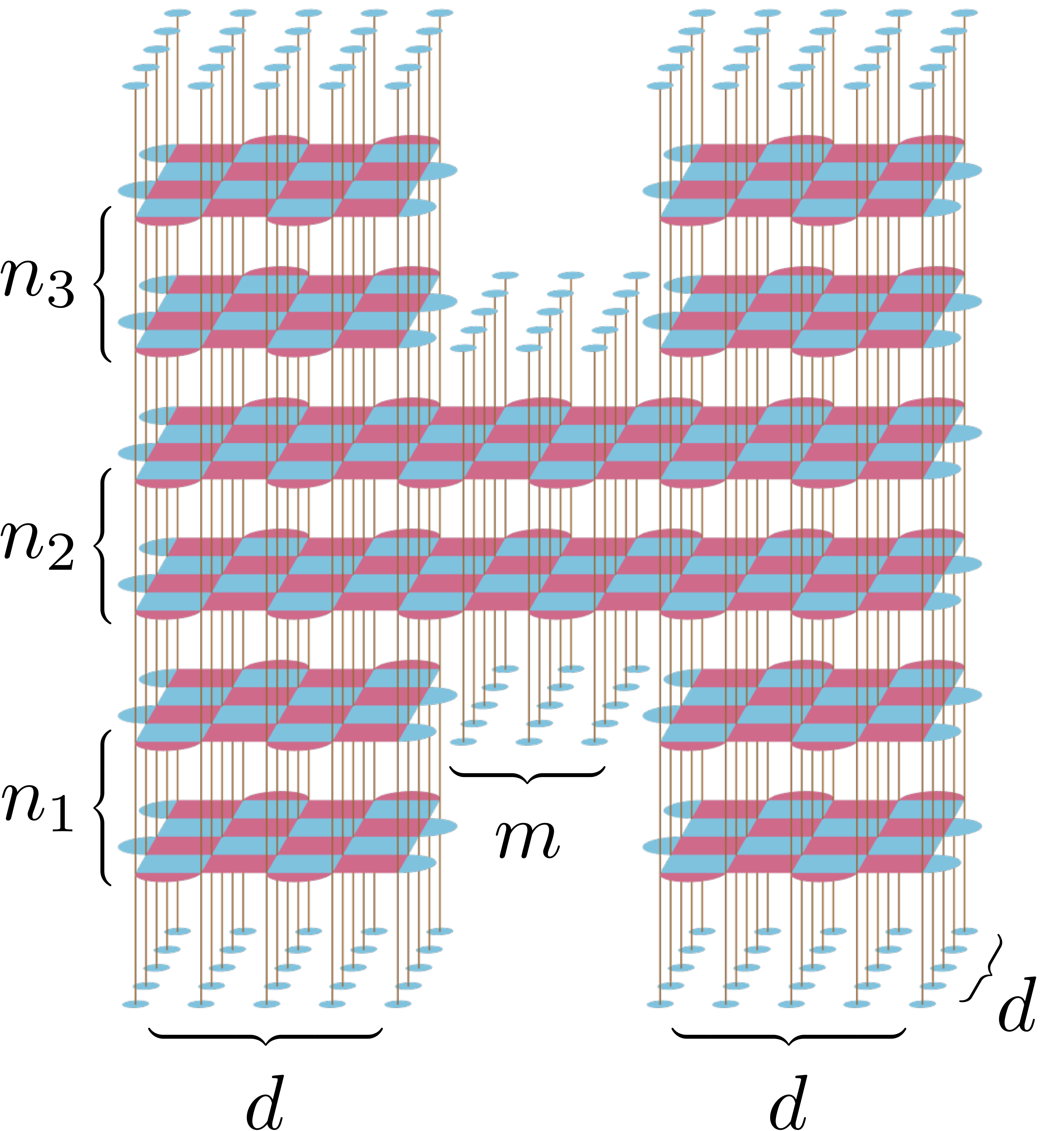}
        \caption{}
    \end{subfigure}
\caption{Illustration of the quantum memory and the Bell-state experiments. A quantum memory experiment on a $d\times d$ lattice initiates $n$ rounds of syndrome extractions. A Bell measurement experiment on two patches of $d\times d$ lattices with routing space length $m$  initiates $n_1$ rounds of syndrome extractions on each patch, then initiates $n_2$ rounds of syndrome extractions on the joint patch by merging the two patches with the routing space, and finally initiates $n_3$ rounds of syndrome extractions on the two split patches. All data qubits are measured under the $Z$-basis before and after their corresponding syndrome extraction cycles.}
\label{fig:latticesurgery}
\end{figure}

Two essential subroutines of the SCQC are the \term{quantum memory experiment} and the \term{Bell-state experiment}. Their quantum circuits are illustrated in \Cref{fig:latticesurgery}. The quantum memory experiment benchmarks the capability of the classical architecture for preserving quantum information, and the Bell-state experiment benchmarks that for essential steps in lattice surgery. As SCQC comprises mostly these two components (in addition to the preparation of a physical magic state and a single-patch logical measurement), we use them to validate our architecture.

Besides real-time execution of quantum circuits with large-scale parallel gates, these SCQC subroutines also require fast processing of classical information in \term{syndrome decoding}. \term{Syndromes} are mid-circuit measurement results indicating errors occurring during the FTQC process. To identify the actual errors and correct them, a dedicated syndrome decoder is needed to deduce the most likely error given the syndrome information. Ideally, the syndrome decoder needs to have a low error rate of inference, and be fast enough in order not to cause exponential syndrome backlog\cite{terhal2015quantum}. Developing and implementing such a low-error, low-latency and high-throughput syndrome decoder is vital to experimental realization of fault-tolerant quantum computation.

\subsection{Validation of scalability}

We first establish the feasibility of our design by implementing an end-to-end prototype quantum computing system, and validating that it functions properly with test qubit calibration programs. In addition, we examine the time variation (\term{jitter}) of pulse control with increasing size of the star-like connection, confirming that our design admits scalable pulse synchronization. A low jitter ensures high-precision synchronization of pulses played on different AWG ports, ensuring high-fidelity controls.

To verify that our design resolves the instruction stress, we execute the aforementioned SCQC subroutines. We profile the running time of the classical controller against the running time of the quantum processor. The classical running time is estimated based on the instruction counts of an in-house CPU profiling tool over the QEMU \mbox{RISC-V} simulator. The running time of the quantum processor is estimated based on previously reported running time of each operation on a comparable superconducting platform\cite{o2017density}. We also quantitatively analyze the cost of instruction decoding and dispatching. Although neither is a scaling-up matter under our architecture design, we quantitatively show that the bandwidth of the differential pairs\cite{T02} can easily afford parallel gate instruction dispatching even under our proof-of-concept ISA implementation.

Real-time classical decoding was previously a hard problem, and even dedicated hardware struggled to achieve real-time decoding for code distance $d$ larger than $11$~\cite{DPM+22,ueno2022neo,battistel2023real}. However, recent advances \cite{TZC+22, Riv22, pymatchingv2} have made real-time decoding much more realistic even on general-purpose CPU. In particular, the sliding-window decoding schemes, introduced independently in \cite{TZC+22} and \cite{Riv22}, parallelize in scale: they split the decoding task evenly into an arbitrary number of parallel threads, with only a small constant overhead factor independent of the number of threads. 
We implement such a parallelized SDU on a \mbox{RISC-V} development board, and benchmark its throughput on increasing code distances. We also give a rough estimation of the bandwidth required for syndrome transmission from the IQE digitizers to the SDU, finding it unlikely to become a bottleneck for our architecture.

\section{System Evaluation}\label{sec:evaluation}

\subsection{Real system demonstration}

\begin{figure}[ht]
\centering
\includegraphics[width=\linewidth]{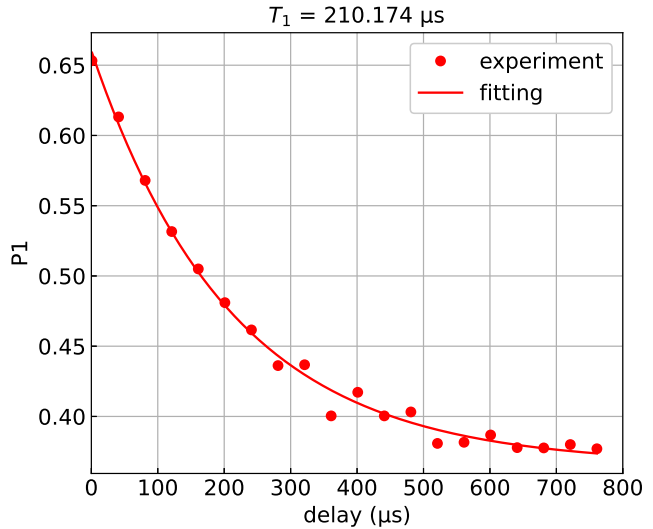}
\caption{T1 measurement on the prototype system.}
\label{fig:T1}
\end{figure}

\TP{Demo through T1 exp}
We implement a prototype system by integrating a \mbox{RISC-V} IP core with our room-temperature electronics, and a demo program to validate the end-to-end quantum computing system consisting of the prototype system, a quantum chip and compilation toolchain. The demo program characterizes a qubit's relaxation time, i.e., the so-called T1 experiment.
We compile an OpenQASM 3.0 front-end code to a \mbox{RISC-V} executable using our in-house compilation toolchain,
and test its correctness both on a pulse-level quantum simulator, and on our in-house superconducting quantum processor. The result of the physical experiment is shown in \Cref{fig:T1}, demonstrating a successful run of the calibration routine.

\subsection{Scalability of maintaining high-fidelity quantum operation}
We now evaluate the feasibility of high-fidelity quantum operations when the chassis-based system is scaled up using a star-like connection.

Skew and jitter, which are crucial for system synchronization, can directly affect the accuracy of quantum operations. Skew, caused by variations in electrical connection length, can usually be compensated for as it remains constant. Jitter, on the other hand, is a greater concern as its effects cannot be calibrated.

To verify the fidelity of quantum operations in a larger system, we set up a 5-layer IQE platform with one MCU, one AWG, and one digitizer in each layer. The main trigger and the root system clock were generated by the MCU in the first layer and transmitted to the MCU in the second layer, and so on for the subsequent layers. 
In each experiment, we pick the first layer and one other layer to test the jitter.  The two AWG output channels from the chosen layers were then connected to a digitizer. We then use fixed-point phase analysis to calculate the jitter between these two signals, as a proxy to evaluate pulse synchronization in larger systems. The critical aspect to consider is whether the jitter varies with the layer distance.

\begin{figure}
    \centering
    \begin{subfigure}[b]{0.2240\textwidth}
        \centering
        \includegraphics[width=\textwidth]{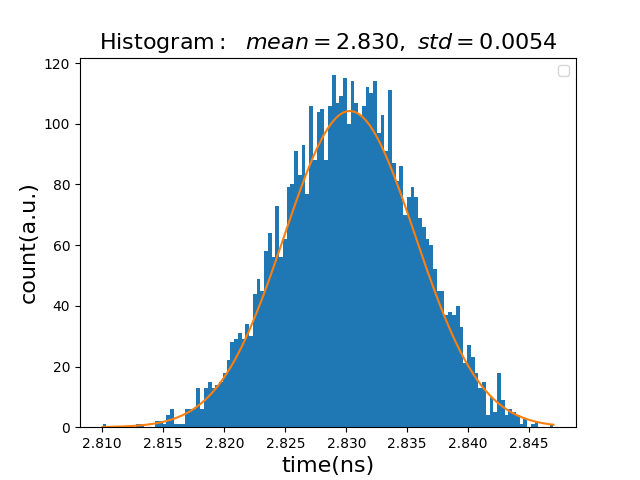}
        \caption {{\small Between layers 1 and 2.}}    
        \label{fig:jitter_1_2}
    \end{subfigure}
    \hfill
    \begin{subfigure}[b]{0.220\textwidth}  
        \centering 
        \includegraphics[width=\textwidth]{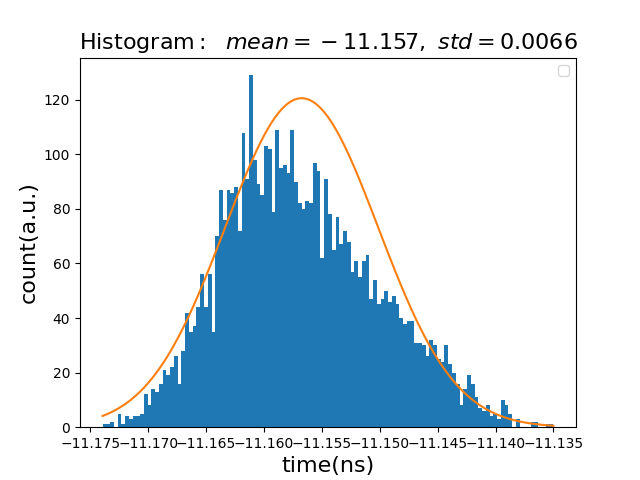}
        \caption[]%
        {{\small Between layers 1 and 3.}}    
        \label{fig:jitter_1_3}
    \end{subfigure}
    \vskip\baselineskip
    \begin{subfigure}[b]{0.220\textwidth}   
        \centering 
        \includegraphics[width=\textwidth]{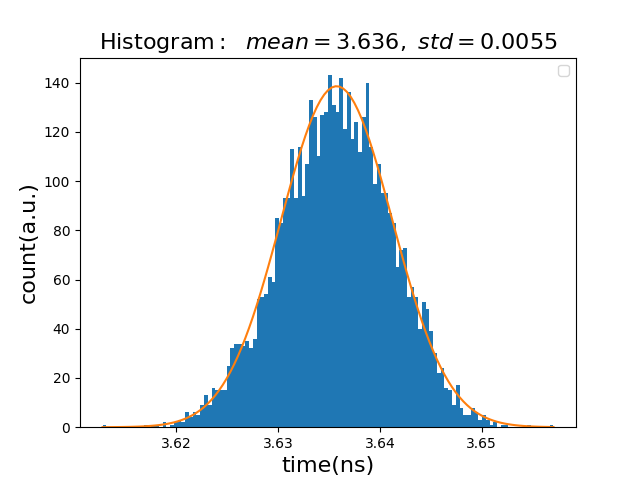}
        \caption[]%
        {{\small Between layers 1 and 4.}}    
        \label{fig:jitter_1_4}
    \end{subfigure}
    \hfill
    \begin{subfigure}[b]{0.220\textwidth}   
        \centering 
        \includegraphics[width=\textwidth]{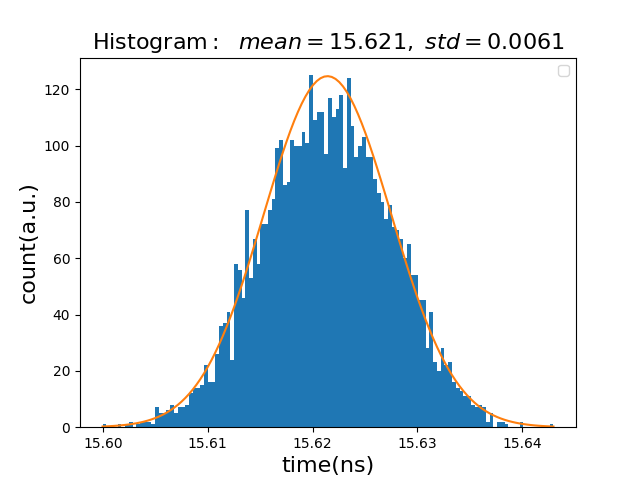}
        \caption[]%
        {{\small Between layers 1 and 5.}}    
        \label{fig:jitter_1_5}
    \end{subfigure}
    \caption{Histogram of the channel-to-channel jitter of two AWGs across chassis in different layers.}
    \label{fig:jitter_histogram}
\end{figure}

As depicted in \Cref{fig:jitter_histogram}, the histograms display the jitter performance at different layer distances. Our measurements of layer-to-layer jitter show that the standard deviation is approximately 6ps, and jitter does not increase with layer distance, indicating effective pulse synchronization within the system.

Based on the $5$-layer results, we conclude that synchronization imprecision of microwave pulses from control electronics across different layers due to phase jitters is minuscule and will not become a major bottleneck for quantum computation.
With a standard chassis that has $18$ slots, the star-like expansion scheme is capable of supporting up to $10^4+10^3+10^2+10+1=11111$ chassis and $111110$ qubits based on the reasonable assumption that a single chassis can drive $10$ additional chassis.

\subsection{Scalability of the instruction pipeline}

\begin{figure}[ht]
    \centering
    \includegraphics[width=\linewidth, trim={0 1.3cm 0 3.6cm}, clip]{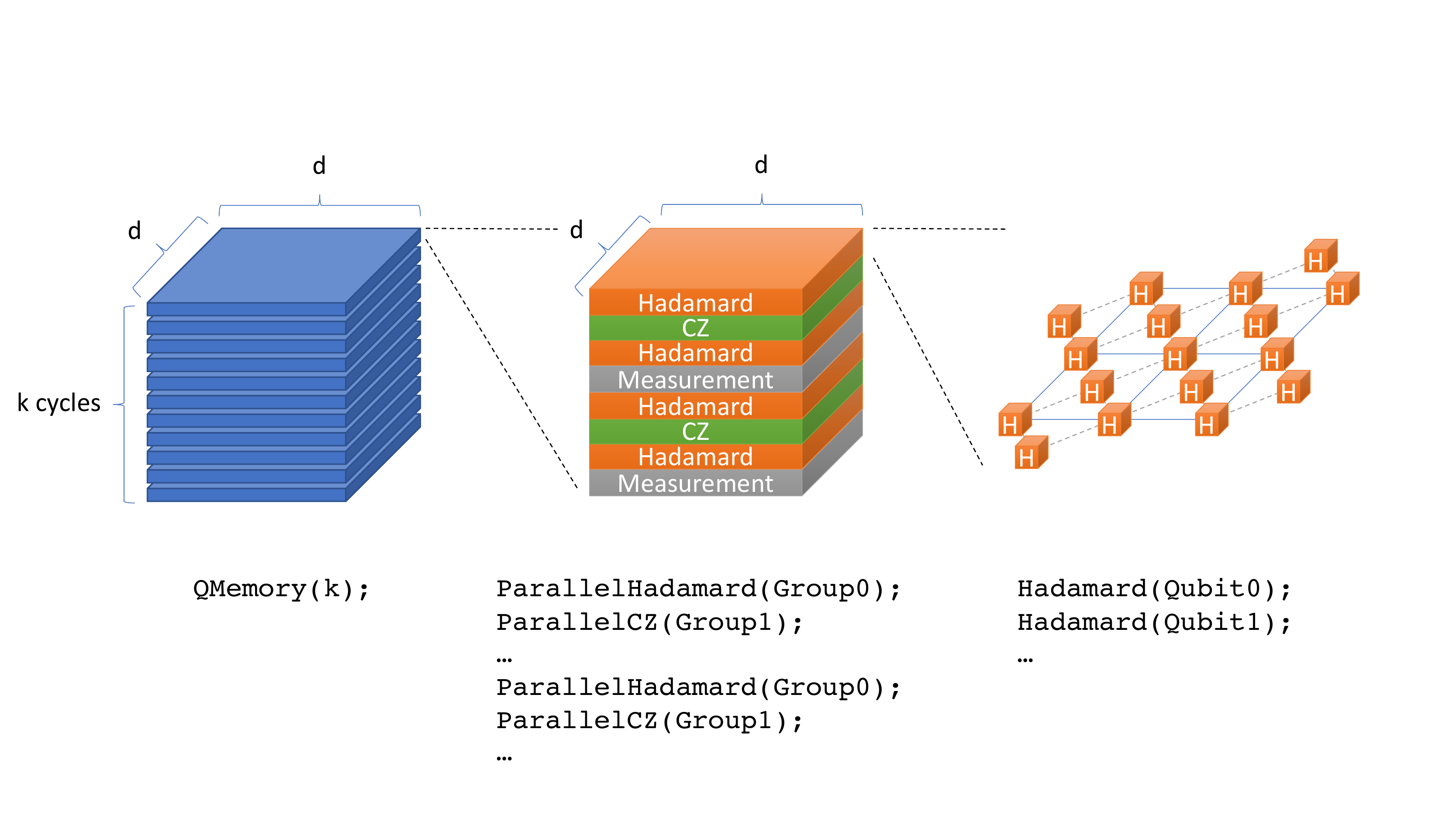}
    \caption{A hierarchical instruction set design for SCQC. The applicability of this hierarchical design is made possible by the MMIO-based infrastructure and the brickwork structure of the syndrome extraction circuits.}
    \label{fig:hierarchy}
\end{figure}

With the MMIO-based custom instruction design, we can test custom CPU-level instructions with different levels of abstraction against the quantum hardware execution time. In particular, we consider the following hierarchy of custom instruction abstraction, illustrated in \Cref{fig:hierarchy}. 

In addition to pulse- and gate-level instructions, the abstraction includes the following instructions.
\begin{itemize}
    \item \textit{Parallel-gate instructions} encode a layer of identical gates acting on a disjoint collection of qubits in a single instruction. Circuits involved in surface code quantum computing have the clear signature of the brickwork structure. Consequently, each round of syndrome extraction only costs a constant number of parallel-gate instructions. In this case, the number of parallel-gate instructions required per unit time no longer scales with the code distance $d$.
    \item \textit{Logical-level instructions} further compresses all operations within a \term{logical cycle}\cite{Lit19a}  into a single instruction. A \term{logical cycle} refers to a repeated structure with $d$ copies of identical syndrome extraction sub-routines, each consisting of constant layers of parallel operations with fixed patterns, with optional single-layer parallel operations before or after the repeats. Such a repeated structure is necessary for fault-tolerance against measurement errors, and serves as building blocks for the SCQC. In this case, the total number of instructions throughout a quantum application stays a constant, regardless of the code distance, leaving more room for improvement when dealing with other scaling factors, such as the number of logical qubits. 
\end{itemize}

\begin{figure}
    \centering
    \begin{subfigure}{0.8\linewidth}
    \includegraphics[width=\linewidth]{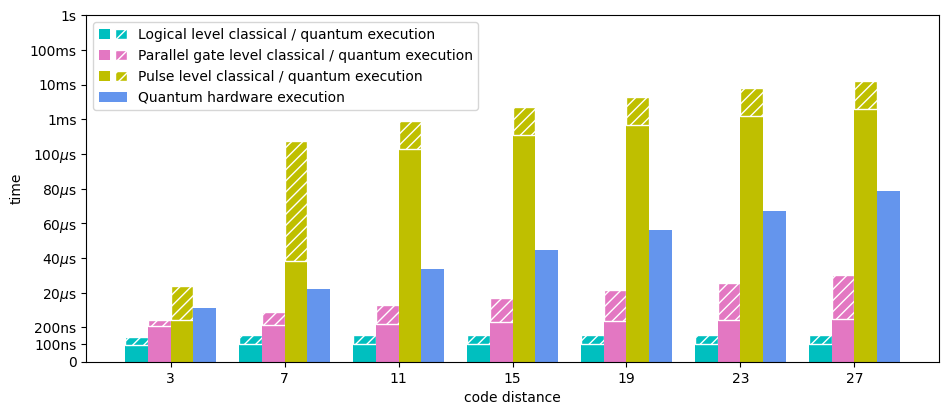}
        \caption{Quantum memory experiment. The number of syndrome extraction rounds is set to $n=\frac{7}{2}(d+1)$.}
    \end{subfigure}
    
    \begin{subfigure}{0.8\linewidth}
    \includegraphics[width=\linewidth]{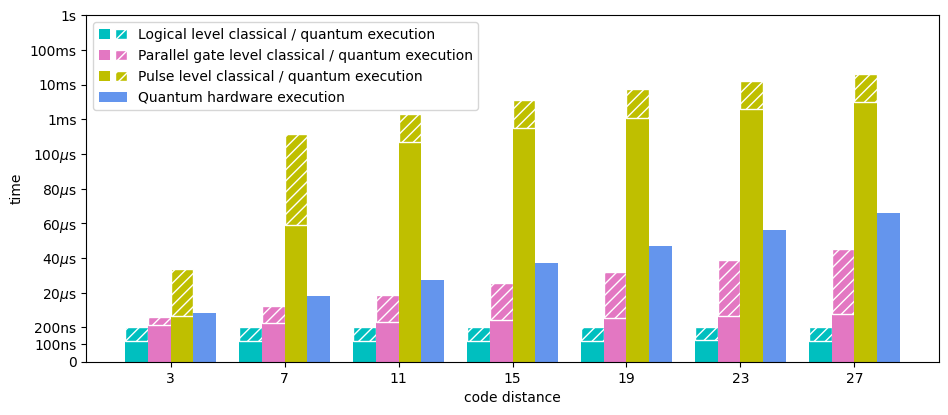}
        \caption{Bell-state experiment. The routing space length is set to $m=3d$, and syndrome extraction rounds are set to $n_1=n_2=n_3=d$.}
        
    \end{subfigure}
    \caption{Comparison of the estimated classical execution time on the MCU against the quantum hardware execution time. The latter is estimated based on $20$ns for each single-qubit gate, $40$ns for each two-qubit gate, and $600$ns for each measurement and reset. The classical run time consists of two parts: 1) For classical instructions, the run time is estimated with the master frequency of the MCU CPU is 1 GHz, and cycle counts from our in-house CPU profiling tools; 2) quantum instructions are executed via expansion into \mbox{RISC-V} base instructions for MMIO, and it takes up to $17$ cycles for each MMIO communication between the MCU and the IQE driver through the system bus. The run time is scaled piece-wise to reflect different running time scaling. The quantum execution time is marked separately with white hatches; note that the proportion of the quantum execution time versus the total classical execution time is distorted owing to the scaling distortion.}
    \label{fig:control_benchmark}
\end{figure}

We estimate the execution time of custom instructions at each abstraction level, scaling in code distance, in \Cref{fig:control_benchmark}. It can be seen that the logical-level instructions stay constant with respect to code distance, and the parallel-gate level instructions scale linearly albeit with a smaller coefficient compared to the quantum running time. The pulse-level instructions scale with $\Theta(d^3)$ and quickly grow into the millisecond region, thus making them infeasible for surface code with reasonable sizes beyond a proof-of-principle demonstration.

For both the memory experiment and the Bell-state experiment, the majority of the quantum execution time is spent on syndrome extraction. Each syndrome cycles takes about 1\textmu s and takes 15 parallel-gate level instructions. This requires a throughput of 0.96 Gbps on the differential pair. This is well below the theoretical limit of the bandwidth of differential pairs\cite{T02}. As this estimation does not scale with respect to the code distance, the transmission of IQE instructions to the control electronics would not become a bottleneck for SCQC.

\subsection{Scalability of the syndrome decoder}
We implement an SDU with a parallel decoding firmware based on the \term{Sandwich Decoder} algorithm~\cite{TZC+22, Riv22}. This firmware splits the decoding task evenly into an arbitrary number of parallel threads, with a small constant overhead factor independent of the thread number, as long as the number of surface code rounds is sufficiently large.

We benchmark its throughput on the aforementioned SCQC subroutines. As a sanity check, we test the SDU implementation on a \mbox{RISC-V} development board, and observe an agreement in results with our QEMU simulator. Benchmarking results are also used to extrapolate the expected throughput if implementing the SDU on other \mbox{RISC-V} IPs. 

In deploying the Sandwich Decoder over the integrated multi-core CPU, the underlying inner decoders are an in-house Union-Find implementation and a recent PyMatching v2\cite{pymatchingv2}. We make several implementation-level improvements on the efficiency for our Union-Find decoder.

To benchmark the performance of our SDU, we apply the Sandwich Decoder to the Bell state experiment; this goes a step further than the memory experiment as in~\cite{TZC+22}. For simplicity, we assume that the routing space between the two logical qubits is small compared to the code distance $d$ and use a single large window to cover the two-qubit measurement part in the overall decoder graph (see \Cref{fig:windows}). All other windows are the same windows used in memory experiments. In our simulation experiments, we input the description of the large window as well as the weight of each edge in that window to the CCU, which randomly generates error syndromes before invoking the decoding module. 

\begin{figure}[!ht]
  \centering
  \begin{subfigure}[t]{0.5\textwidth}
  \centering
\begin{tikzpicture}[scale=1.4]
  \begin{groupplot}[
      group style={
          group name=my plots,
          group size=1 by 1,
          xlabels at=edge bottom,
          xticklabels at=edge bottom,
          vertical sep=0pt
      },
      legend style={
        font=\tiny,
        /tikz/every even column/.append style={column sep=0.02cm}
            },  
      scale=1,
      minor xtick={7,9,11,13,17,19,21,23,27,29,31,33,37,39,41,43,47,49,51,53,57,59,61,63,67},
      xtick={5,15,25,35,45,55,65},
      xlabel={code distance $d$},
      xmin=0,xmax=70,
      tickpos=left
  ]
  \nextgroupplot[ymode=log, ylabel={decoding throughput / \textmu s},minor ytick={0.5,0.6,0.7,0.8,0.9,1,2,3,4,5,6,7,8,9,10,20,30,40,50,60,70,80,90,100,200,300,400,500,600,700,800,900,1000,2000,3000}, ymin=0.3,ymax=3000, ylabel absolute, y label style={at={(axis description cs:0.1,0.5)}}, legend style={at={(0.72,0.5)},anchor=north, font=\tiny,nodes={scale=0.8, transform shape}}, label style={font=\tiny},tick label style={font=\tiny}]
  \addplot[blue,mark=*,mark options={scale=0.5}] plot coordinates {
    (5,41/3/5)
    (7,181/4/5)
    (9,425/5/5)
    (11,856/6/5)
    (13,1223/7/5)
    (15,2113/8/5)
    (17,3083/9/5)
    (19,3853/10/5)
    (21,5743/11/5)
    (23,7483/12/5)
    (25,9534/13/5)
    (27,10814/14/5)
    (29,14814/15/5)
    (31,16375/16/5)
    (33,21950/17/5)
    (35,26014/18/5)
    (37,27857/19/5)
    (39,35554/20/5)
    (41,41432/21/5)
    (43,47664/22/5)
    (45,54612/23/5)
    (47,61890/24/5)
    (49,69492/25/5)
    (51,78718/26/5)
    (53,87865/27/5)
    (55,98508/28/5)
    (57,108455/29/5)
    (59,121320/30/5)
    (61,134088/31/5)
    (63,147232/32/5)
    (65,160551/33/5)
    (67,175785/34/5)
  };
  \addlegendentry{SDU+in-house UF, p=0.0001}
  \addplot[cyan,mark=*,mark options={scale=0.5}] plot coordinates {
    (5,54/3/5)
    (7,321/4/5)
    (9,783/5/5)
    (11,1546/6/5)
    (13,2500/7/5)
    (15,4120/8/5)
    (17,5688/9/5)
    (19,7908/10/5)
    (21,11276/11/5)
    (23,14332/12/5)
    (25,18488/13/5)
    (27,21988/14/5)
    (29,28571/15/5)
    (31,33706/16/5)
    (33,42839/17/5)
    (35,50063/18/5)
    (37,56417/19/5)
    (39,68872/20/5)
    (41,78951/21/5)
    (43,93475/22/5)
    (45,107093/23/5)
    (47,119622/24/5)
    (49,136165/25/5)
  };
  \addlegendentry{SDU+in-house UF, p=0.001}  
  \addplot[color=olive,mark=*,mark options={scale=0.5}]
    plot coordinates {
    (5,104/3/5)
    (7,788/4/5)
    (9,2276/5/5)
    (11,4681/6/5)
    (13,7971/7/5)
    (15,12851/8/5)
    (17,18668/9/5)
    (19,26499/10/5)
    (21,37616/11/5)
    (23,47814/12/5)
    (25,63095/13/5)
    (27,79696/14/5)};
  \addlegendentry{SDU+in-house UF, p=0.005}
  
  \addplot[dashed,blue,mark=o,mark options={scale=0.5, solid}] plot coordinates {
    (5,10/3/5)
    (7,29/4/5)
    (9,67/5/5)
    (11,115/6/5)
    (13,188/7/5)
    (15,273/8/5)
    (17,388/9/5)
    (19,543/10/5)
    (21,705/11/5)
    (23,958/12/5)
    (25,1423/13/5)
    (27,1714/14/5)
    (29,2308/15/5)
    (31,2832/16/5)
    (33,3447/17/5)
    (35,4252/18/5)
    (37,5097/19/5)
    (39,5966/20/5)
    (41,7016/21/5)
    (43,8954/22/5)
    (45,10294/23/5)
    (47,11736/24/5)
    (49,12852/25/5)
    (51,14838/26/5)
    (53,15797/27/5)
    (55,17095/28/5)
    (57,19502/29/5)
  };
  \addlegendentry{SDU+PyMatching2\cite{pymatchingv2}, p=0.0001}
  \addplot[dashed,cyan,mark=o,mark options={scale=0.5, solid}] plot coordinates {
    (5,74/3/5)
    (7,218/4/5)
    (9,494/5/5)
    (11,1000/6/5)
    (13,1776/7/5)
    (15,2878/8/5)
    (17,4433/9/5)
    (19,6359/10/5)
    (21,8668/11/5)
    (23,12290/12/5)
    (25,15938/13/5)
    (27,19710/14/5)
    (29,24525/15/5)
    (31,29408/16/5)
    (33,35889/17/5)
    (35,42730/18/5)
    (37,50051/19/5)
    (39,58755/20/5)
    (41,75632/21/5)
    (43,86170/22/5)
    (45,97905/23/5)
    (47,113397/24/5)
    (49,125224/25/5)
    (51,141622/26/5)
  };
  \addlegendentry{SDU+PyMatching2\cite{pymatchingv2}, p=0.001}  
  \addplot[dashed,color=olive,mark=o,mark options={scale=0.5, solid}]
    plot coordinates {
    (5,365/3/5)
    (7,1214/4/5)
    (9,3324/5/5)
    (11,6540/6/5)
    (13,11498/7/5)
    (15,17744/8/5)
    (17,27729/9/5)
    (19,40403/10/5)
    (21,54872/11/5)
    (23,72951/12/5)
    (25,96816/13/5)
    (27,120570/14/5)};
  \addlegendentry{SDU+PyMatching2\cite{pymatchingv2}, p=0.005}
  
  \addplot[dotted, red, thick] plot coordinates {
    (0,40)
    (70,40)
  };
\addlegendentry{SiFive P650\cite{SiFive22}/T-head C910\cite{CXL+20}}
  \addplot[densely dotted, red, thick] plot coordinates {
    (0,1088)
    (70,1088)
  };
\addlegendentry{ET-SoC-1\cite{DEA+21}}
  \end{groupplot}
  \end{tikzpicture}
  \caption{Decoding throughput for the quantum memory experiment.}
  \end{subfigure}
  
   \par\bigskip
   
  \begin{subfigure}[t]{0.5\textwidth}
  \centering
\begin{tikzpicture}[scale=1.4]
  \begin{groupplot}[
      group style={
          group name=my plots,
          group size=1 by 1,
          xlabels at=edge bottom,
          xticklabels at=edge bottom,
          vertical sep=0pt
      },
      legend style={
        font=\tiny,
        /tikz/every even column/.append style={column sep=0.02cm}
            },  
      scale=1,
     height=5cm,
      minor xtick={7,9,11,13,17,19,21,23,27,29,31,33,37,39,41,43,47,49,51,53,57,59,61,63,67},
      xtick={5,15,25,35,45,55,65},
      xlabel={code distance $d$},
      xmin=0,xmax=70,
      tickpos=left
  ]
  \nextgroupplot[ymode=log, minor ytick={20,30,40,50,60,70,80,90,100,200,300,400,500,600,700,800,900,1000,2000,3000}, ymin=10,ymax=3000, ylabel={decoding throughput / \textmu s}, ylabel absolute, y label style={at={(axis description cs:0.1,0.5)}}, legend style={at={(0.72,0.5)},anchor=north, font=\tiny,nodes={scale=0.8, transform shape}}, label style={font=\tiny},tick label style={font=\tiny}]
  \addplot[blue,mark=*,mark options={scale=0.5}] plot coordinates {
	(9,2*950/5/5)
	(11,2*1616/6/5)
	(13,2*2673/7/5)
	(15,2*3977/8/5)
	(17,2*5459/9/5)
	(19,2*7666/10/5)
	(21,2*10035/11/5)
	(23,2*13482/12/5)
	(25,2*16742/13/5)
	(27,2*21441/14/5)
	(29,2*25786/15/5)
	(31,2*32088/16/5)
	(33,2*37714/17/5)
	(35,2*47044/18/5)
  };
  \addlegendentry{SDU+in-house UF, p=0.0001}
  \addplot[cyan,mark=*,mark options={scale=0.5}] plot coordinates {
	(9,2*1588/5/5)
	(11,2*3020/6/5)
	(13,2*5002/7/5)
	(15,2*7540/8/5)
	(17,2*10522/9/5)
	(19,2*14694/10/5)
	(21,2*19150/11/5)
	(23,2*25801/12/5)
	(25,2*32504/13/5)
	(27,2*41271/14/5)
  };
  \addlegendentry{SDU+in-house UF, p=0.001}  
  \addplot[color=olive,mark=*,mark options={scale=0.5}]
    plot coordinates {
	(9,2*5131/5/5)
	(11,2*9723/6/5)
	(13,2*16486/7/5)
	(15,2*25763/8/5)
    };
  \addlegendentry{SDU+in-house UF, p=0.005}
  \addplot[dashed,blue,mark=o,mark options={scale=0.5,solid}] plot coordinates {
	(9,2*161/5/5)
	(11,2*276/6/5)
	(13,2*454/7/5)
	(15,2*703/8/5)
	(17,2*1058/9/5)
	(19,2*1555/10/5)
	(21,2*2119/11/5)
	(23,2*2896/12/5)
	(25,2*3843/13/5)
	(27,2*4970/14/5)
	(29,2*6255/15/5)
	(31,2*7724/16/5)
	(33,2*9453/17/5)
	(35,2*11279/18/5)
	(37,2*13452/19/5)
	(39,2*15822/20/5)
	(41,2*18360/21/5)
	(43,2*21835/22/5)
	(45,2*24925/23/5)
	(47,2*28393/24/5)
	(49,2*32185/25/5)
	(51,2*35865/26/5)
	(53,2*40668/27/5)
	(55,2*45710/28/5)
	(57,2*50526/29/5)
	(59,2*56385/30/5)
	(61,2*62179/31/5)
	(63,2*68601/32/5)
	(65,2*75056/33/5)
  };
  \addlegendentry{SDU+PyMatching2\cite{pymatchingv2}, p=0.0001}
  \addplot[dashed,cyan,mark=o,mark options={scale=0.5,solid}] plot coordinates {
	(9,2*1441/5/5)
	(11,2*2933/6/5)
	(13,2*5142/7/5)
	(15,2*8228/8/5)
	(17,2*12137/9/5)
	(19,2*17101/10/5)
	(21,2*22660/11/5)
	(23,2*30091/12/5)
	(25,2*38897/13/5)
  };
  \addlegendentry{SDU+PyMatching2\cite{pymatchingv2}, p=0.001}  
  \addplot[dashed,color=olive,mark=o,mark options={scale=0.5,solid}]
    plot coordinates {
	(9,2*9484/5/5)
	(11,2*18789/6/5)
	(13,2*32672/7/5)
    };
  \addlegendentry{SDU+PyMatching2\cite{pymatchingv2}, p=0.005}

  \end{groupplot}
  \end{tikzpicture}
  \caption{Decoding throughput for the Bell-state experiment.}
  \end{subfigure}
  \caption{The average syndrome decoding throughput for the quantum memory experiment and the two-qubit joint measurement in the Bell state experiment. Decoding throughput is defined as the average processing time per single layer of syndrome on a single core with 1GHz master frequency. We experimentally benchmark the total running time of both experiments under different code distances and multiple runs, and deduce the per-layer running time. For the ease of comparison, we set the step size as $t+1 = (d+1)/2$ and the window size as $3(t+1)$ for both experiments. 
  For the quantum memory experiments, the red dashed lines indicate the capability of the specific \mbox{RISC-V} SoCs, converted to the same scale as the experimental data. The data points below a certain dashed line indicate the feasibility of running the corresponding task on the corresponding SoC within 1\textmu s. 
}
  \label{fig:benchmark}
\end{figure}
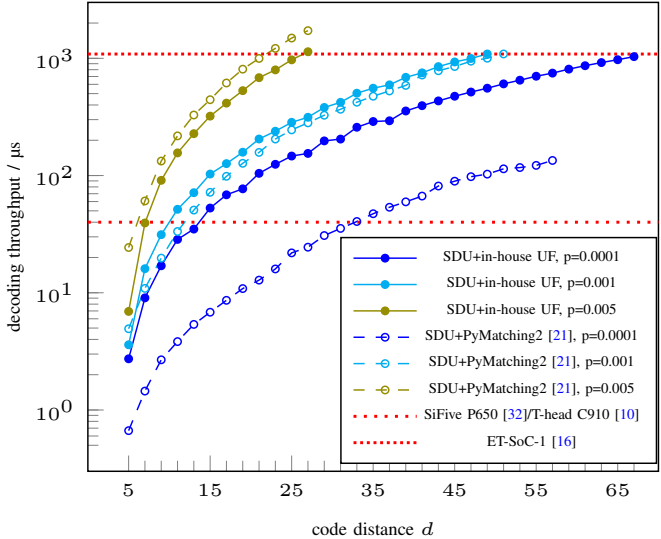
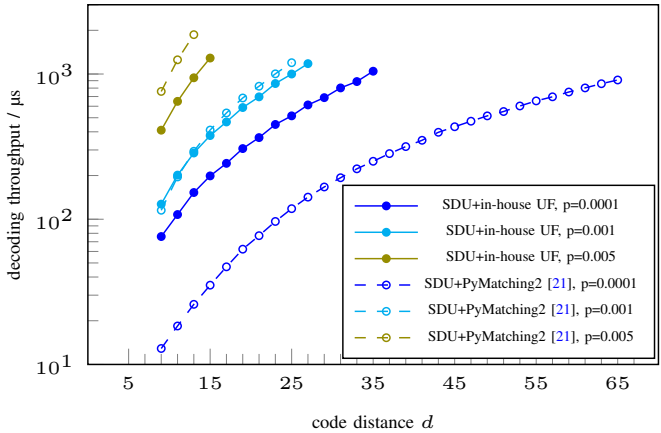

We use the benchmarking results on the development board, shown in~\Cref{fig:benchmark}, to estimate the decoding time when implementing our SDU on various \mbox{RISC-V} SoCs. With Union-Find and PyMatching 2 as inner decoders, the implemented SDU can decode up to distances $13$ and $31$ on SiFive P650~\cite{SiFive22}, T-head C910, or comparable alternatives~\cite{CXL+20} (16 cores at 2.5GHz), or $67$
and $57$ with ET-SoC-1~\cite{DEA+21}(1088 cores at 1GHz), all within just 1 microsecond for physical error rate $p=0.0001$. Our evaluation of PyMatching 2 shows that its performance is constrained by the limited 1GB memory available on the tested development board. We expect that PyMatching 2 could achieve even better results on a higher-end development board with a larger memory.

Besides the decoding throughput, another constraint for the decoder architecture is that the large amount of syndrome information generated throughout the SCQC process must not  saturate the communication bandwidth.
It is known that raw syndrome measurement results can quickly become a bandwidth bottleneck~\cite{DPM+22}, thus must be compressed. One approach is to record the \term{detection events}, i.e., changes in a sequence of syndrome bits, rather than all the syndrome bits.  For a quantum memory experiment with a code distance $d$ and a syndrome extraction cycles $n$, each ancilla qubit needs to generate $p_\text{detect}\cdot n\log_2n$ bits on average, assuming that each detection event happens with a probability $p_\text{detect}$. Roughly, the bandwidth requirement would become 100 Mbps for $p_\text{detect}=0.02$ and $n=d=33$. This compression can be done on each digitizer separately before transmission to the IQE driver. More advanced compression algorithms may achieve a better compression rate, but may require conjoined processing from different digitizers. Such an algorithm can be placed in the IQE driver should there be a bottleneck in the MMIO bandwidth.

\begin{figure}
    \centering
    \begin{tikzpicture}[join=round, scale=0.90]
    \newcommand{\coord}[3]{(#1+#3*0.707, #2+#3*0.707)}
    \tikzstyle{coreface} = [draw=orange,fill=orange!20,fill opacity=0.5]
    \tikzstyle{ann} = [fill=white,font=\footnotesize,inner sep=1pt]
    \newcommand{\boxwindow}[2]{
        \draw\coord{0}{0}{1}--\coord{0}{1}{1}--\coord{1}{1}{1}--\coord{1}{0}{1}--cycle;
        \draw\coord{0}{0}{0}--\coord{0}{0}{1}--\coord{0}{1}{1}--\coord{0}{1}{0}--cycle;
        \draw\coord{0}{0}{0}--\coord{0}{0}{1}--\coord{1}{0}{1}--\coord{1}{0}{0}--cycle;
        \filldraw[coreface]\coord{0}{#1}{1}--\coord{0}{#2}{1}--\coord{1}{#2}{1}--\coord{1}{#1}{1}--cycle;
        \filldraw[coreface]\coord{0}{#1}{0}--\coord{0}{#1}{1}--\coord{0}{#2}{1}--\coord{0}{#2}{0}--cycle;
        \filldraw[coreface]\coord{0}{#1}{0}--\coord{0}{#1}{1}--\coord{1}{#1}{1}--\coord{1}{#1}{0}--cycle;
        \filldraw[coreface]\coord{1}{#1}{0}--\coord{1}{#1}{1}--\coord{1}{#2}{1}--\coord{1}{#2}{0}--cycle;
        \filldraw[coreface]\coord{0}{#2}{0}--\coord{0}{#2}{1}--\coord{1}{#2}{1}--\coord{1}{#2}{0}--cycle;
        \filldraw[coreface]\coord{0}{#1}{0}--\coord{0}{#2}{0}--\coord{1}{#2}{0}--\coord{1}{#1}{0}--cycle;
        \draw\coord{1}{0}{0}--\coord{1}{0}{1}--\coord{1}{1}{1}--\coord{1}{1}{0}--cycle;
        \draw\coord{0}{1}{0}--\coord{0}{1}{1}--\coord{1}{1}{1}--\coord{1}{1}{0}--cycle;
        \draw\coord{0}{0}{0}--\coord{0}{1}{0}--\coord{1}{1}{0}--\coord{1}{0}{0}--cycle;
    }
    \begin{scope}[]
    \boxwindow{0}{0.6}
    \end{scope}
    \begin{scope}[xshift=2cm]
    \boxwindow{0}{0.6}
    \end{scope}
    \begin{scope}[yshift=1.8cm]
    
    \draw\coord{0}{0}{1}--\coord{0}{2.6}{1}--\coord{1}{2.6}{1}--\coord{1}{1.8}{1}--\coord{2}{1.8}{1}--\coord{2}{2.6}{1}--\coord{3}{2.6}{1}--\coord{3}{0}{1}--\coord{2}{0}{1}--\coord{2}{0.8}{1}--\coord{1}{0.8}{1}--\coord{1}{0}{1}--cycle;
    \filldraw[coreface]\coord{0}{0.4}{1}--\coord{0}{2.2}{1}--\coord{1}{2.2}{1}--\coord{1}{1.8}{1}--\coord{2}{1.8}{1}--\coord{2}{2.2}{1}--\coord{3}{2.2}{1}--\coord{3}{0.4}{1}--\coord{2}{0.4}{1}--\coord{2}{0.8}{1}--\coord{1}{0.8}{1}--\coord{1}{0.4}{1}--cycle;
    
    \draw\coord{0}{0}{0}--\coord{0}{0}{1}--\coord{0}{2.6}{1}--\coord{0}{2.6}{0}--cycle;
    \filldraw[coreface]\coord{0}{0.4}{0}--\coord{0}{0.4}{1}--\coord{0}{2.2}{1}--\coord{0}{2.2}{0}--cycle;
    
    \draw\coord{0}{0}{0}--\coord{0}{0}{1}--\coord{1}{0}{1}--\coord{1}{0}{0}--cycle;
    \filldraw[coreface]\coord{0}{0.4}{0}--\coord{0}{0.4}{1}--\coord{1}{0.4}{1}--\coord{1}{0.4}{0}--cycle;
    
    \filldraw[coreface]\coord{1}{0.4}{0}--\coord{1}{0.4}{1}--\coord{1}{0.8}{1}--\coord{1}{0.8}{0}--cycle;
    \draw\coord{1}{0}{0}--\coord{1}{0}{1}--\coord{1}{0.8}{1}--\coord{1}{0.8}{0}--cycle;
    
    \draw\coord{1}{0.8}{0}--\coord{1}{0.8}{1}--\coord{2}{0.8}{1}--\coord{2}{0.8}{0}--cycle;
    \filldraw[coreface]\coord{1}{0.8}{0}--\coord{1}{0.8}{1}--\coord{2}{0.8}{1}--\coord{2}{0.8}{0}--cycle;
    
    \draw\coord{2}{0}{0}--\coord{2}{0}{1}--\coord{2}{0.8}{1}--\coord{2}{0.8}{0}--cycle;
    \filldraw[coreface]\coord{2}{0.4}{0}--\coord{2}{0.4}{1}--\coord{2}{0.8}{1}--\coord{2}{0.8}{0}--cycle;
    
    \draw\coord{2}{0}{0}--\coord{2}{0}{1}--\coord{3}{0}{1}--\coord{3}{0}{0}--cycle;
    \filldraw[coreface]\coord{2}{0.4}{0}--\coord{2}{0.4}{1}--\coord{3}{0.4}{1}--\coord{3}{0.4}{0}--cycle;
        
    \filldraw[coreface]\coord{0}{2.2}{0}--\coord{0}{2.2}{1}--\coord{1}{2.2}{1}--\coord{1}{2.2}{0}--cycle;
    \draw\coord{0}{2.6}{0}--\coord{0}{2.6}{1}--\coord{1}{2.6}{1}--\coord{1}{2.6}{0}--cycle;
    
    \filldraw[coreface]\coord{1}{1.8}{0}--\coord{1}{1.8}{1}--\coord{1}{2.2}{1}--\coord{1}{2.2}{0}--cycle;
    \draw\coord{1}{1.8}{0}--\coord{1}{1.8}{1}--\coord{1}{2.6}{1}--\coord{1}{2.6}{0}--cycle;
    
    \filldraw[coreface]\coord{1}{1.8}{0}--\coord{1}{1.8}{1}--\coord{2}{1.8}{1}--\coord{2}{1.8}{0}--cycle;
    \draw\coord{1}{1.8}{0}--\coord{1}{1.8}{1}--\coord{2}{1.8}{1}--\coord{2}{1.8}{0}--cycle;

    \draw\coord{2}{1.8}{0}--\coord{2}{1.8}{1}--\coord{2}{2.6}{1}--\coord{2}{2.6}{0}--cycle;
    \filldraw[coreface]\coord{2}{1.8}{0}--\coord{2}{1.8}{1}--\coord{2}{2.2}{1}--\coord{2}{2.2}{0}--cycle;
    
    \filldraw[coreface]\coord{2}{2.2}{0}--\coord{2}{2.2}{1}--\coord{3}{2.2}{1}--\coord{3}{2.2}{0}--cycle;
    \draw\coord{2}{2.6}{0}--\coord{2}{2.6}{1}--\coord{3}{2.6}{1}--\coord{3}{2.6}{0}--cycle;
    
    \filldraw[coreface]\coord{3}{0.4}{0}--\coord{3}{0.4}{1}--\coord{3}{2.2}{1}--\coord{3}{2.2}{0}--cycle;
    \draw\coord{3}{0}{0}--\coord{3}{0}{1}--\coord{3}{2.6}{1}--\coord{3}{2.6}{0}--cycle;
    
    \filldraw[coreface]\coord{0}{0.4}{0}--\coord{0}{2.2}{0}--\coord{1}{2.2}{0}--\coord{1}{1.8}{0}--\coord{2}{1.8}{0}--\coord{2}{2.2}{0}--\coord{3}{2.2}{0}--\coord{3}{0.4}{0}--\coord{2}{0.4}{0}--\coord{2}{0.8}{0}--\coord{1}{0.8}{0}--\coord{1}{0.4}{0}--cycle;
    \draw\coord{0}{0}{0}--\coord{0}{2.6}{0}--\coord{1}{2.6}{0}--\coord{1}{1.8}{0}--\coord{2}{1.8}{0}--\coord{2}{2.6}{0}--\coord{3}{2.6}{0}--\coord{3}{0}{0}--\coord{2}{0}{0}--\coord{2}{0.8}{0}--\coord{1}{0.8}{0}--\coord{1}{0}{0}--cycle;
    
    \end{scope}
    \begin{scope}[yshift=5.2cm]
    \boxwindow{0.4}{1}
    \end{scope}
    \begin{scope}[xshift=2cm, yshift=5.2cm]
    \boxwindow{0.4}{1}
    \end{scope}
\end{tikzpicture}
    \caption{Illustration of a division of the overall three-dimensional decoder graph of the Bell-state experiment (\Cref{fig:latticesurgery}) into windows. Under the assumption of a small routing space, even though the center window is large, its size is still $O(d)\times O(d)\times O(d)$. Note that the center window covers more layers than other windows; an alternative scheme (not depicted) is to divide the center window further so that every window covers the same number of layers, which gives rise to more complicated windows. 
    }
    \label{fig:windows}
\end{figure}
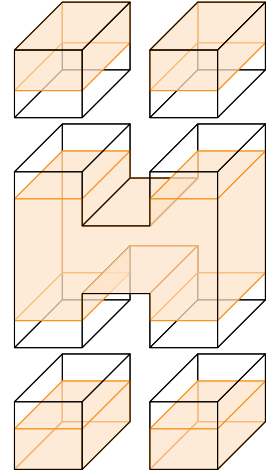

\section{Discussion and outlook}
We present a scalable design for the classical architecture of quantum computing. Our design aims towards easy scaling with no significant overhead. We evaluate its scalability on two basic subroutines over a prominent fault-tolerant scheme, and validate its practical feasibility with a prototype implementation.

A natural next step is to implement the real system with quantum processors of a much larger scale than that in our study. The current design is estimated to scale up easily over thousands of qubits. Such an estimation is based on the size of the allocated MMIO addresses, the picosecond accuracy in the synchronization of the trigger signal across different electronics, and the physical size of the electronics stacks. Although most of the limiting factors can be lifted through a more careful design, it remains uncertain if unforeseen problems may arise with a larger-scale quantum processor. A possible further scaling-up through modularization is to let each MCU control one or a few logical components, such as a single logical qubit or a patch of the routing space, and let an upper-level control unit issue logical instructions to these logical components while maintaining synchronization. 

In this study, we assume room-temperature devices for their wide adoption at the time of writing. However, our design in principle is not limited to such, and may in particular work well for cryogenic electronics, such as cryo-CMOS\cite{charbon2016cryo} or single-flux-quantum\cite{mukhanov2011energy}, as long as the component functionalities can be implemented. Such demonstrations would be an interesting future direction. 

Another important direction is to demonstrate through more sophisticated tasks than our two ``toy-model'' subroutines. Such experiments may lead to the discovery of currently unknown limiting factors for classical architecture in SCQC.

Classical architecture is just half of the story, as numerous challenges still to be addressed in quantum architecture. Beyond the quantum processor's scale, hurdles such as input/output (I/O) management, interconnection, packaging, and heat and power dissipation must be overcome. Previous research in quantum architecture has often more focused on the feasibility of qubit control than the potential demands of intensive classical computation. Conversely, studies on classical architecture have primarily examined the viability of specific classical computation tasks such as syndrome decoding, either in-fridge or out-of-fridge, under the bold assumption that high-fidelity qubit control can be realistically achieved. A holistic evaluation of the FTQC workflow, encompassing both classical and quantum architectures, will aid in identifying potential bottlenecks and determing the most effective steps to move forward. 

\section*{Acknowledgements}
We would like to thank all the members of DAMO Quantum Laboratory who contributed to the development of the quantum hardware used to demonstrate the end-to-end workflow in this study. This work was supported by Alibaba Group through Alibaba Research Intern Program, and conducted when Yihuai Gao and Liwei Qiu were research interns at Alibaba Group.

\bibliographystyle{IEEEtranS}


\end{document}